\documentclass[aps,prd,preprintnumbers,twocolumn,groupedaddress,nofootinbib]{revtex4}
\pdfoutput=1
\usepackage{graphicx}
\usepackage{latexsym}
\def\beq{\begin{equation}}
\def\eeq{\end{equation}}
\def\bey{\begin{eqnarray}}
\def\eey{\end{eqnarray}}

\def\lsim{\mathrel{\raise.3ex\hbox{$<$\kern-.75em\lower1ex\hbox{$\sim$}}}}
\def\gsim{\mathrel{\raise.3ex\hbox{$>$\kern-.75em\lower1ex\hbox{$\sim$}}}}

\newcommand{\be}{\begin{equation}}
\newcommand{\ee}{\end{equation}}
\newcommand{\tev}{\ensuremath{\mathrm{\,Te\kern -0.1em V}}\xspace}
\newcommand{\gev}{\ensuremath{\mathrm{\,Ge\kern -0.1em V}}\xspace}
\newcommand{\tevt}{\ensuremath{\mathrm{Te\kern -0.1em V}}\xspace}
\newcommand{\kev}{\ensuremath{\mathrm{\,ke\kern -0.1em V}}\xspace}
\newcommand{\mev}{\ensuremath{\mathrm{\,Me\kern -0.1em V}}\xspace}

\begin{document}

\title{The Empirical Case For 10 GeV Dark Matter}
\author{Dan Hooper}
\affiliation{Center for Particle Astrophysics, Fermi National Accelerator Laboratory, Batavia, IL 60510, USA}
\affiliation{Department of Astronomy and Astrophysics, University of Chicago, Chicago, IL 60637, USA}
\affiliation{Kavli Institute for Cosmological Physics, University of Chicago, Chicago, IL 60637 USA}

\date{\today}

\begin{abstract}

In this article, I summarize and discuss the body of evidence which has accumulated in favor of dark matter in the form of approximately 10 GeV particles. This evidence includes the spectrum and angular distribution of gamma rays from the Galactic Center, the synchrotron emission from the Milky Way's radio filaments, the diffuse synchrotron emission from the Inner Galaxy (the ``WMAP Haze'') and low-energy signals from the direct detection experiments DAMA/LIBRA, CoGeNT and CRESST-II. This collection of observations can be explained by a relatively light dark matter particle with an annihilation cross section consistent with that predicted for a simple thermal relic ($\sigma v \sim 10^{-26}$ cm$^3$/s) and with a distribution in the halo of the Milky Way consistent with that predicted from simulations. Astrophysical explanations for the gamma ray and synchrotron signals, in contrast, have not been successful in accommodating these observations. Similarly, the phase of the annual modulation observed by DAMA/LIBRA (and now supported by CoGeNT) is inconsistent with all known or postulated modulating backgrounds, but are in good agreement with expectations for dark matter scattering. This scenario is consistent with all existing indirect and collider constraints, as well as the constraints placed by CDMS. Consistency with xenon-based experiments can be achieved if the response of liquid xenon to very low-energy nuclear recoils is somewhat suppressed relative to previous evaluations, or if the dark matter possesses different couplings to protons and neutrons.

\end{abstract}


\maketitle

\section{Introduction}

Several independent lines of observational evidence support the conclusion that the majority of the matter in our universe consists of cold dark matter, rather than baryons or other known particle species~\cite{hep-ph/0404175,bullet}. These observations, however, reveal little about the nature of the dark matter itself. An enormous variety of dark matter candidates have been proposed, ranging in mass from $\sim$$10^{-6}$ eV axions to superheavy ({\it i.e.}~GUT or Planck scale) particles. From among this vast landscape of dark matter candidates, the class known as weakly interacting massive particles (WIMPs) are among the most strongly motivated. The hierarchy problem requires new physics to appear at or around the electroweak scale, but in order to be consistent with the stringent constraints of electroweak precision measurements, the interactions of those particles must be limited, such as by a symmetry or parity which in many cases leads to the stability of one or more state. A stable particle with a weak-scale mass, $X$, will be produced and freeze-out in the early universe with a thermal relic density given by $\Omega_{X}  h^2 \approx 0.1 \times [\sigma v/(3\times 10^{-26} \, {\rm cm}^3/{\rm s})]^{-1}$, where $\sigma v$ is the self-annihilation cross section of the particle, evaluated at the time of freeze-out. As $\sigma v\sim 3\times 10^{-26}$ cm$^3$/s is similar to the value estimated for a generic weak-scale interaction, we conclude that a GeV-TeV scale stable particle with a roughly weak-scale annihilation cross section will naturally be produced in the early universe with an abundance similar to the observed density of dark matter. This argument, sometimes referred to as the ``WIMP Miracle'', applies equally well to particles with 1-20 GeV masses as to those with masses more traditionally associated with supersymmetric neutralino dark matter ($m_{\chi}\sim 40-1000$ GeV). It is not at all difficult to construct a viable particile physics model which includes a $\sim$10 GeV WIMP that is produced in the early universe with an abundance equal to the observed density of dark matter.

If the dark matter consists of WIMPs, these particles could potentially be observed through a variety of techniques. Direct detection experiments attempt to observe the recoil from the elastic scattering of dark matter particles interacting with nuclei in a detector. Indirect detection experiments are designed to observe and identify the annihilation products of WIMPs, such as gamma-rays, neutrinos, cosmic rays, and emission at radio/microwave wavelengths. Alternatively, one could potentially produce and observe dark matter particles in collider experiments, such as at the Large Hadron Collider (LHC). While each of these approaches have their advantages and disadvantages, it is interesting to note that all three of these strategies for detecting dark matter particles have reached or are about to reach the level of sensitivity that has long anticipated to be required to observe most postulated varieties of WIMPs. 


Over the past several years, a number of observational signals have been reported which can be interpreted as interactions of dark matter particles. While anomalous or otherwise difficult to explain astrophysical signals are often interpreted as possible products of dark matter annihilations (for example, Refs.~\cite{astro-ph/0309686,hep-ph/0309029,astro-ph/0405235,arXiv:0810.4995,803110,Dobler:2011mk}), these anomalies are in most cases ultimately found to have non-exotic origins, whether astrophysical or instrumental. In order for the scientific community to become convinced that a given signal or collection of signals does in fact arise from dark matter particles, those observations will have to be favorably compared to the predictions of the dark matter hypothesis in several different ways. Ideally, the set of observations will overconstrain the problem in such a way that conclusions can be made which are largely independent of astrophysical uncertainties and choices in the particle dark matter model.

In this article, I will attempt to make the case that these stringent criteria required to convincingly identify dark matter interactions are largely satisfied by the body of evidence that has accumulated in favor of $\sim$10 GeV dark matter particles. These data include the spectral and morphological distribution of gamma-rays from the Galactic Center~\cite{hooperlinden,HG2}, the synchrotron emission from the Inner Galaxy~\cite{timhaze,darkhaze,darkhaze1}, the synchrotron emission from radio filaments in the Inner Galaxy~\cite{filaments}, and signals from three direct detection experiments, DAMA/LIBRA~\cite{damanew}, CoGeNT~\cite{Aalseth:2011wp,Aalseth:2010vx}, and CRESST-II~\cite{Angloher:2011uu}. As I will describe in more detail later this in this article, the gamma-ray signal observed from the Galactic Center is consistent with 7-12 GeV dark matter particles annihilating mostly to leptons with an annihilation cross section consistent with that of a thermal relic (as motivated by the ``WIMP Miracle'', $\sigma v\sim 3 \times 10^{-26}$ cm$^3$/s), and with a distribution in good agreement with the results of hydrodynamical simulations ($\rho_{\rm DM}\propto r^{-1.3}$, where $r$ is the distance to the Galactic Center). No viable astrophysical explanations for this emission are known (see Ref.~\cite{hooperlinden} for a discussion). Using this choice for the dark matter mass, annihilation cross section, annihilation channels, and spatial distribution, one can predict the spectrum, intensity, and angular distribution of synchrotron emission resulting from electron and positron dark matter annihilation products in the Inner Galaxy and compare this to that observed by WMAP (the ``WMAP Haze'') and from the Milky Way's radio filaments.  In each of these cases, there is good agreement between these observations and the predictions of the gamma-ray motivated dark matter model. This scenario is further supported by the observations reported by the DAMA/LIBRA~\cite{damanew}, CoGeNT~\cite{Aalseth:2011wp,Aalseth:2010vx}, and CRESST~\cite{Angloher:2011uu} collaborations, which each report signals consistent with a dark matter particle of similar mass. These three experiments make use of different technologies, target materials, and detection strategies, but each report results which are not compatible with known backgrounds, but that can be accommodated by a dark matter particle with a $\sim$10 GeV mass and an elastic scattering cross section with nuclei of approximately a few times $10^{-41}$ cm$^2$~\cite{arXiv:1110.5338}.

These six distinct observations (of the Fermi Galactic Center, the Milky Way radio filaments, the WMAP Haze, and signals from DAMA/LIBRA, CoGeNT and CRESST-II) provide a collection of evidence for $\sim$10 GeV dark matter particles with 1) is unable to be explained by proposed or known backgrounds, 2) overconstrains the properties of the underlying dark matter model, and 3) is consistent with theoretical expectations. By overconstraining the model, I mean that multiple observations require the same distribution, mass, and cross sections for dark matter. For example, one could not interpret the spectra of the Milky Way's radio filaments as signals of dark matter annihilations if the WMAP Haze was not also observed -- the annihilation rate and channels required to power the radio filaments requires a Haze-like signal to also be present. Similarly, if the radial profile of the WMAP Haze or the collection of radio filaments had a much shallower or steeper distribution, it would not be easily reconciled with the dark matter profile implied by the gamma ray observations of the Galactic Center. Furthermore, if CoGeNT had seen no evidence of annual modulation in their event rate, it would be very difficult to interpret DAMA/LIBRA's modulation as dark matter, and vice versa. By consistent with theoretical expectations, I mean that the dark matter particle required to explain these observations possesses an annihilation cross section consistent with a simple thermal relic, and does not require any unexpected or baroque features (such as large boost factors, non-standard dark matter distributions, or non-minimal particle physics features such as Sommerfeld enhancements or inelastic scattering).

The primary purpose of this article is to summarize in a self-consistent way these observations and their implications for dark matter. Much of the material described here has been presented previously elsewhere, and the reader is encouraged to follow the references (in particular Refs.~\cite{hooperlinden,filaments,timhaze,arXiv:1110.5338,Buckley:2010ve}) to find many of the details that have been omitted here. 

The remainder of the article is structured as follows. In Sec.~\ref{indirect}, I discuss the gamma-ray signal from the Galactic Center as observed by the Fermi Gamma-Ray Space Telescope, as well as the synchrotron signals observed from the Milky Way's radio filaments, and from the Inner Galaxy by WMAP. In Sec.~\ref{direct}, I discuss the direct detection signals reported by DAMA/LIBRA, CoGeNT, and CRESST-II. In Sec.~\ref{particle}, I discuss the particle physics implications of these observations and explore some of features of models that contain a dark matter candidate capable of producing these signals. Finally, in Sec.~\ref{summary}, I summarize these results and draw conclusions.

\section{Evidence From Indirect Detection}
\label{indirect}

\subsection{Expectations and Predictions}

\subsubsection{General Comments}

Before discussing any specific observations, I will begin by asking the question, ``What would a 10 GeV annihilating dark matter particle look like to indirect detection experiments?'' Although the answer to this question depends to a degree on the detailed properties of the dark matter particle being considered, a few very general and model-independent statements can be made. One the one hand, as the total dark matter annihilation rate is proportional to $1/m^2_{\rm DM}$, dark matter particles with relatively light masses are expected to produce significantly brighter annihilation signals than are predicted from heavier particles. On the other hand, many indirect detection experiments have energy thresholds which make them insensitive to the annihilation products of low mass dark matter particles. Large volume neutrino telescopes such as IceCube (DeepCore), for example, are sensitive only to neutrinos above $\sim$50-100 GeV ($\sim$10 GeV). For light dark matter particles, we are thus forced to rely on much smaller neutrino detectors with lower energy thresholds, such as Super-Kamiokande. Similarly, ground-based gamma-ray telescopes such as HESS, VERITAS and MAGIC are almost entirely blind to gamma-rays below $\sim$50-100 GeV. Furthermore, the spectrum of $\sim$1-10 GeV cosmic rays is significantly impacted by the effects of the solar winds, diffusive reacceleration, convection, and other astrophysical phenomena which make them more difficult to model and interpret than their higher energy counterparts. 

So while many indirect detection experiments are not sensitive to relatively light dark matter particles, those which are able to detect low-energy annihilation products are likely to observe quite large fluxes of such products. Particularly promising strategies for identifying such light dark matter particles are those being employed by the Fermi Gamma-Ray Space Telescope (FGST) and various radio and microwave telescopes. For $\sim$10 GeV dark matter particles with an annihilation cross sections on the order of $\sigma v\sim 3\times 10^{-26}$ cm$^3$/s, these experiments are generally predicted to observe quite bright and likely observable signals.

\subsubsection{The Annihilation Rate in the Inner Galaxy}

The annihilation rate of dark matter at a given location in space depends on both the annihilation cross section of the particle and on the square of its number density. And while we do not know precisely how the dark matter is distributed or with what cross section it annihilates, we do possess information which enables us to make reasonable and informed estimates of these quantities. 

As stated in the introduction, the dark matter annihilation cross section is related to its thermal relic abundance. In particular, a stable particle with a mass in the GeV to TeV range will be produced thermally in the early universe with a density equal to the measured dark matter abundance if it self-annihilates with a cross section of $\sigma v\approx 3\times 10^{-26}$ cm$^3$/s~\cite{kolbturner}. For a GeV-TeV thermal relic, this can be thought of as an approximate upper limit on the annihilation cross section today (unless very light force carriers lead to Sommerfeld enhancements~\cite{sommerfeld}). It is possible that the annihilation cross section today could be lower than this value if velocity-dependent terms in the annihilation cross section contribute significantly to the process of thermal freeze-out, but do not contribute significantly to the current annihilation rate. Coannihilations between the dark matter and another state could also play an important role in freeze out~\cite{threeexceptions}, although for the light mass range being considered here, this is unlikely to be the case. Taken together, the relic abundance calculation leads us to expect the dark matter to annihilate with a cross section as large as, and likely not very much smaller than, $\sigma v \sim 3\times 10^{-26}$ cm$^3$/s.

Our knowledge of the distribution of dark matter in the Milky Way is based on a combination of observational constraints and numerical simulations. Observations of the Milky Way's rotation curve and its gravitational microlensing optical depth are best fit by a cusped dark matter distribution, $\rho_{\rm DM} \propto r^{-1.3}$, although with large uncertainties~\cite{bertonehalo}. And while numerical simulations which model the evolution of cold dark matter without baryons tend to find halos with inner profiles of approximately $\rho_{\rm DM} \propto r^{-1}$~\cite{nfw,vialactea,aquarius}, hydrodynamical simulations which include the baryonic processes involved in galaxy formation have begun to converge in favor of Milky Way-like halos being significantly contracted~\cite{ac}, leading to a steepening of their inner profiles from $\rho_{\rm DM} \propto r^{-1}$ to slopes typically in the range of $\rho_{\rm DM} \propto r^{-1.2}$ to $r^{-1.5}$ (see Ref.~\cite{Gnedin:2011uj} and references therein).


The highest annihilation rates occur in the high density central regions of dark matter halos. The center of the Milky Way, in particular, has long been recognized as the single most promising target of indirect detection efforts~\cite{gc}. 10 GeV dark matter particles annihilating with a cross section of $\sigma v = 3\times 10^{-26}$ cm$^3$/s and distributed as $\rho_{\rm DM} \propto r^{-1.3}$ release energy in annihilation products at a rate of $\sim 2 \times 10^{40}$ GeV/s within the innermost 150 parsecs around the Galactic Center (corresponding to approximately the innermost $1^{\circ}$). As we will see in the remainder of this section, this power is comparable to that observed in gamma-rays from the Galactic Center by Fermi. This annihilation rate is also in good agreement with that required to power the observed synchrotron emission from the Milky Way's radio filaments and the synchrotron emission from the inner galaxy known as the ``WMAP Haze''.

\begin{figure*}[t]
\centering
\includegraphics[angle=0.0,width=2.45in]{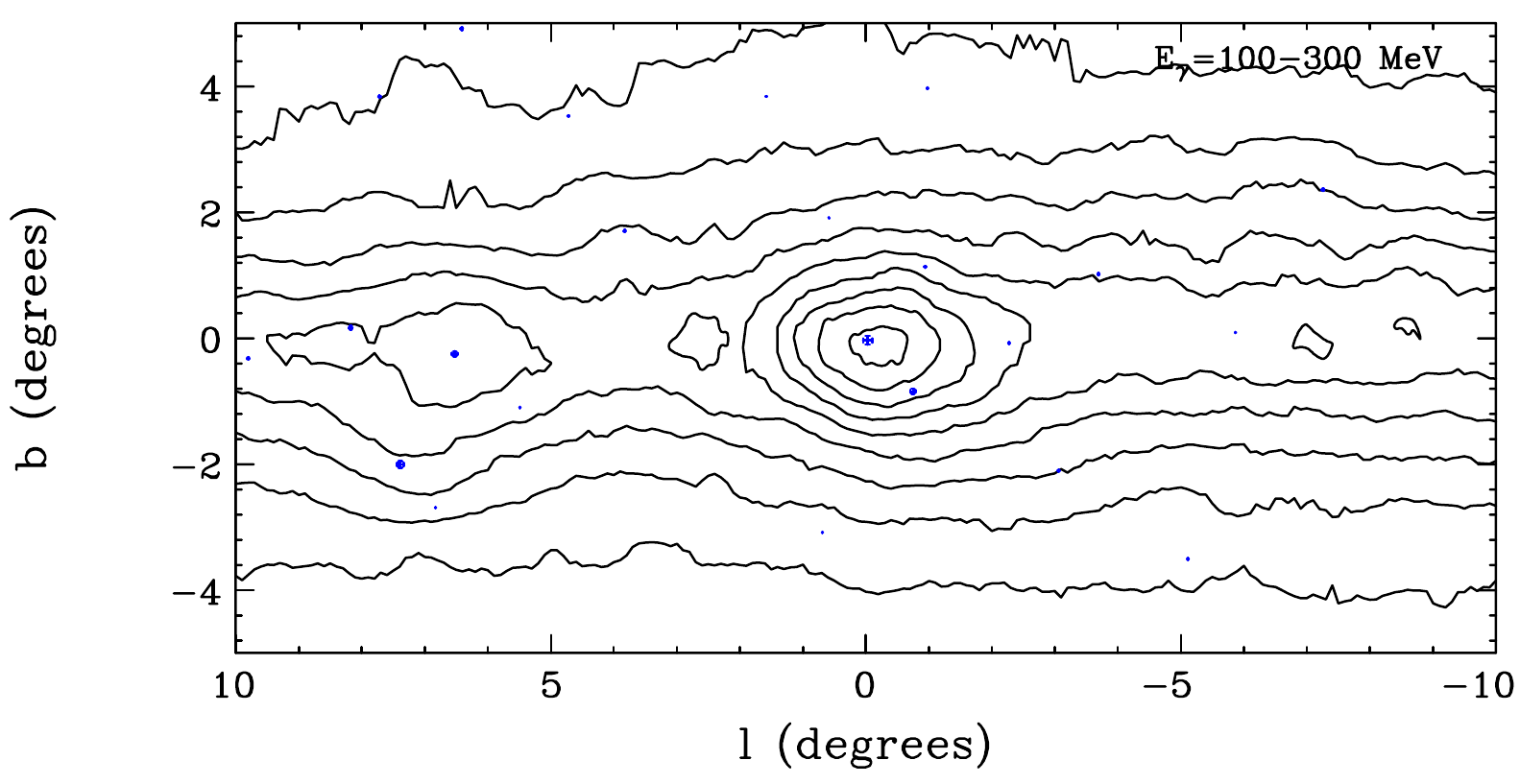}
\includegraphics[angle=0.0,width=2.2in]{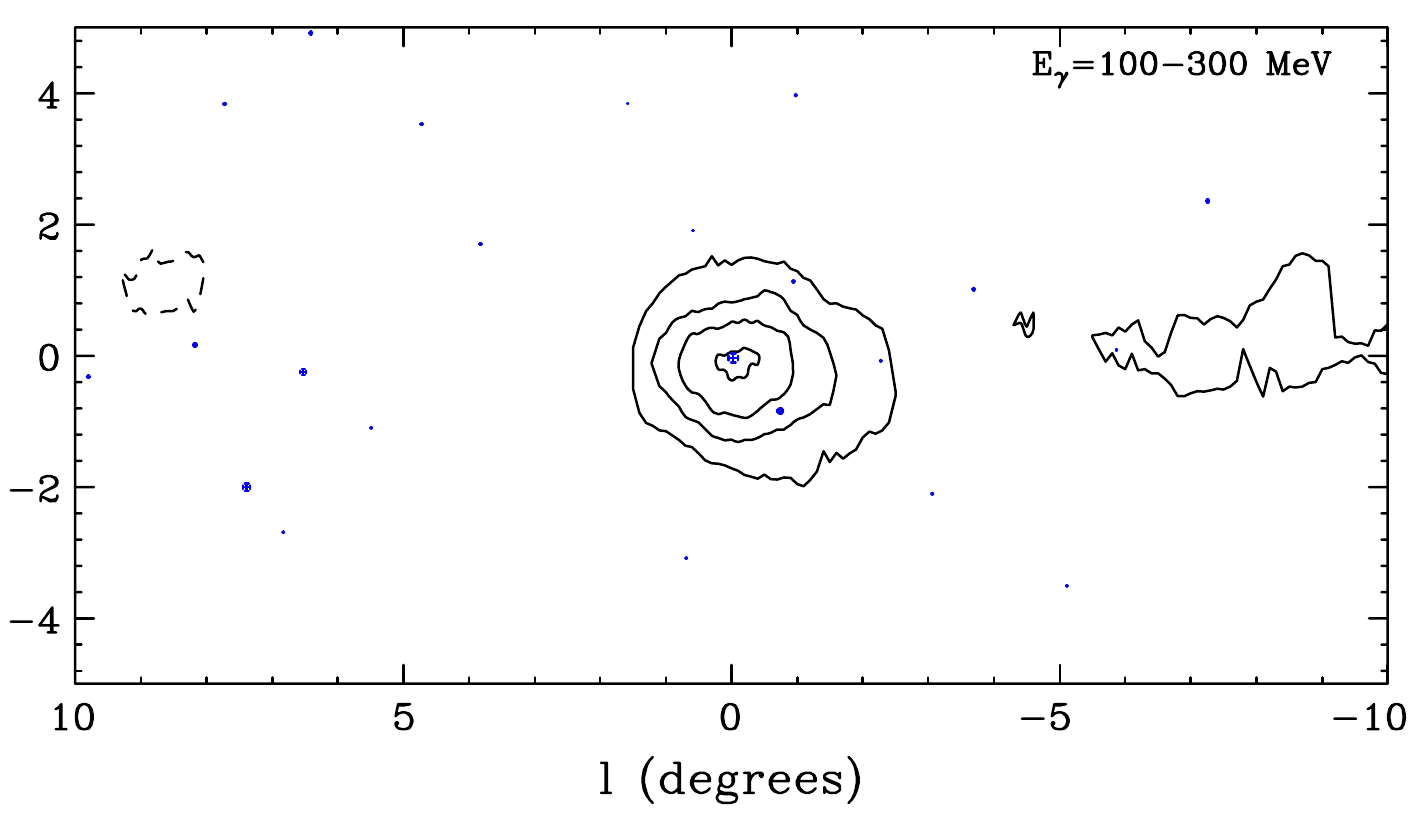}\\
\includegraphics[angle=0.0,width=2.45in]{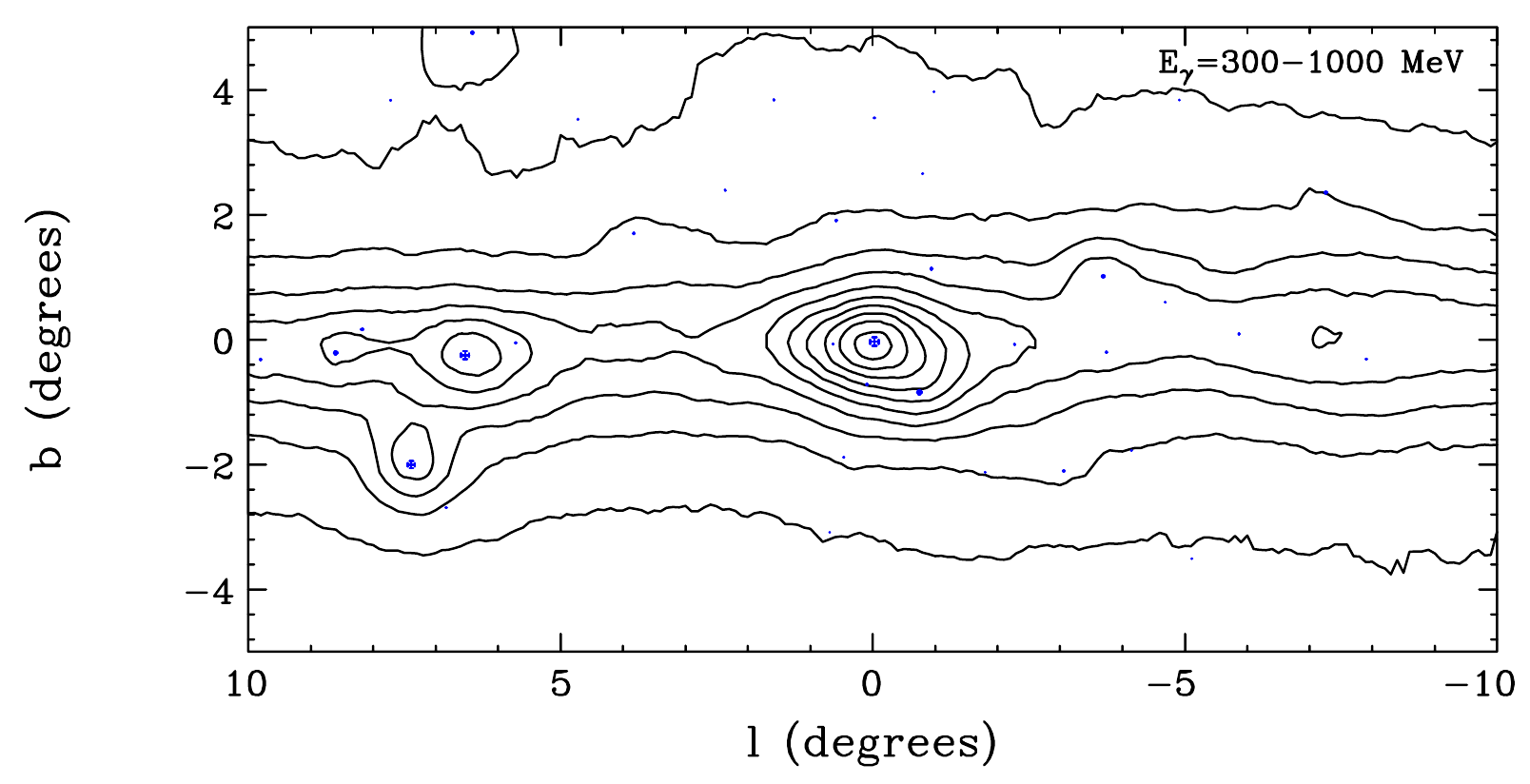}
\includegraphics[angle=0.0,width=2.2in]{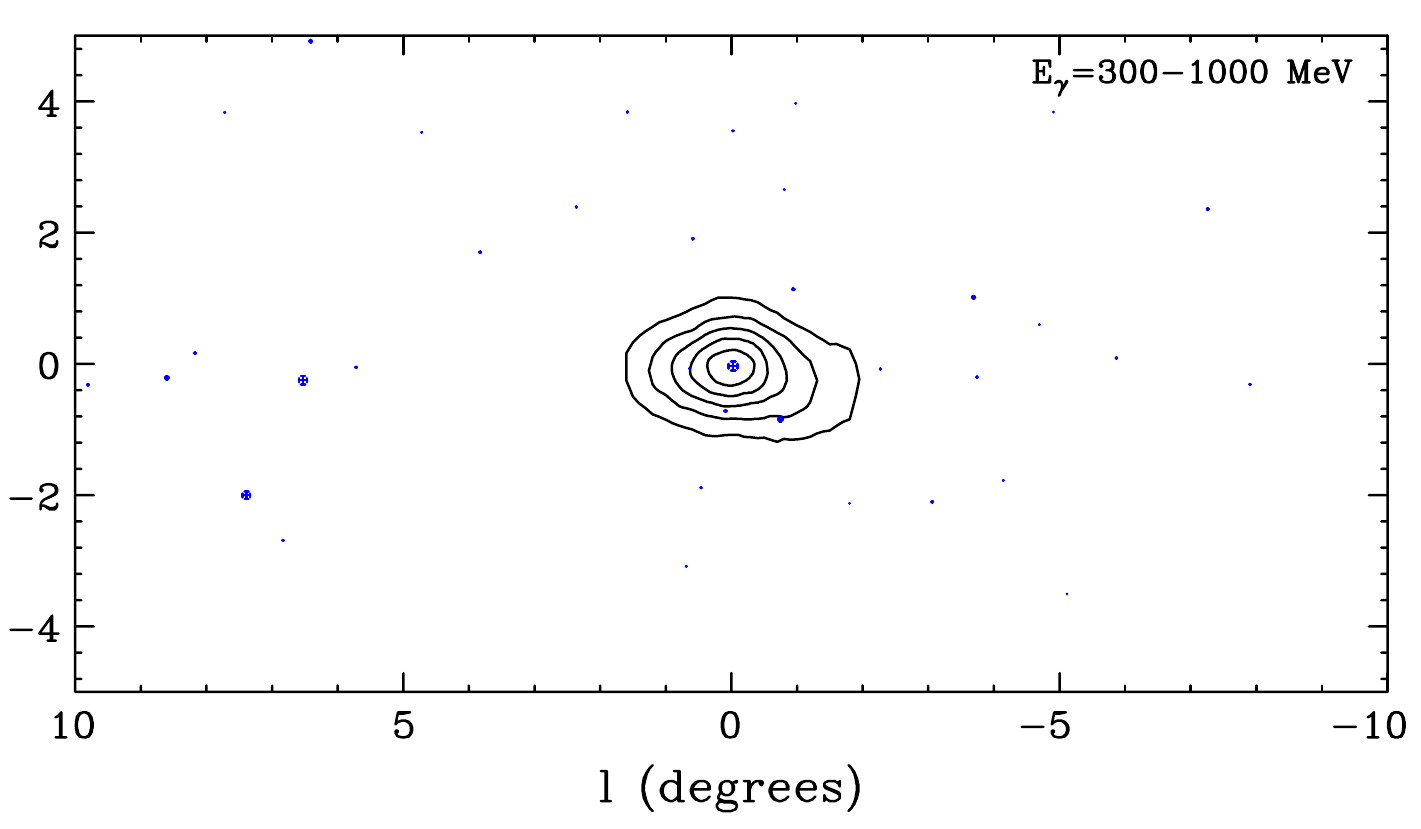}\\
\includegraphics[angle=0.0,width=2.45in]{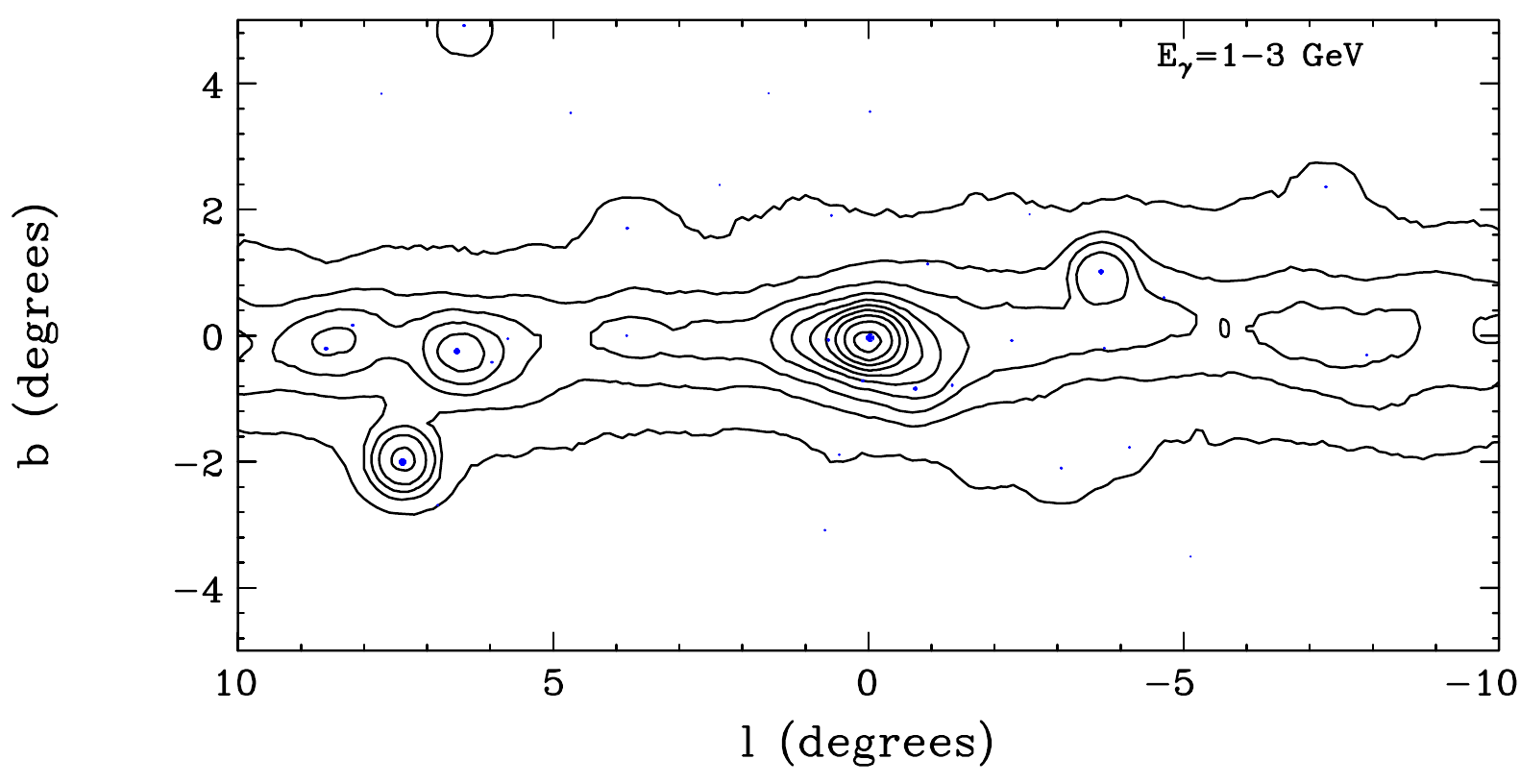}
\includegraphics[angle=0.0,width=2.2in]{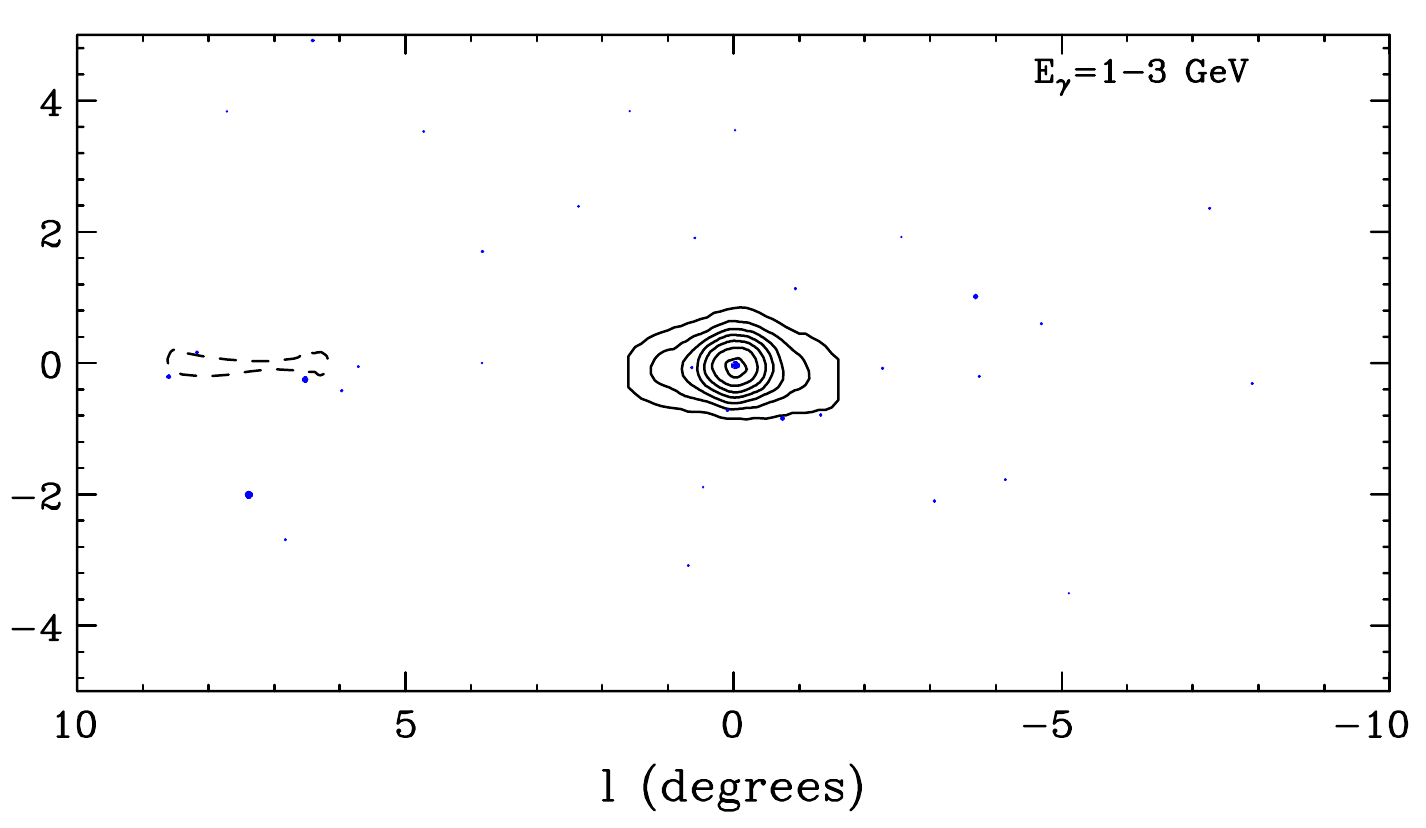}\\
\includegraphics[angle=0.0,width=2.45in]{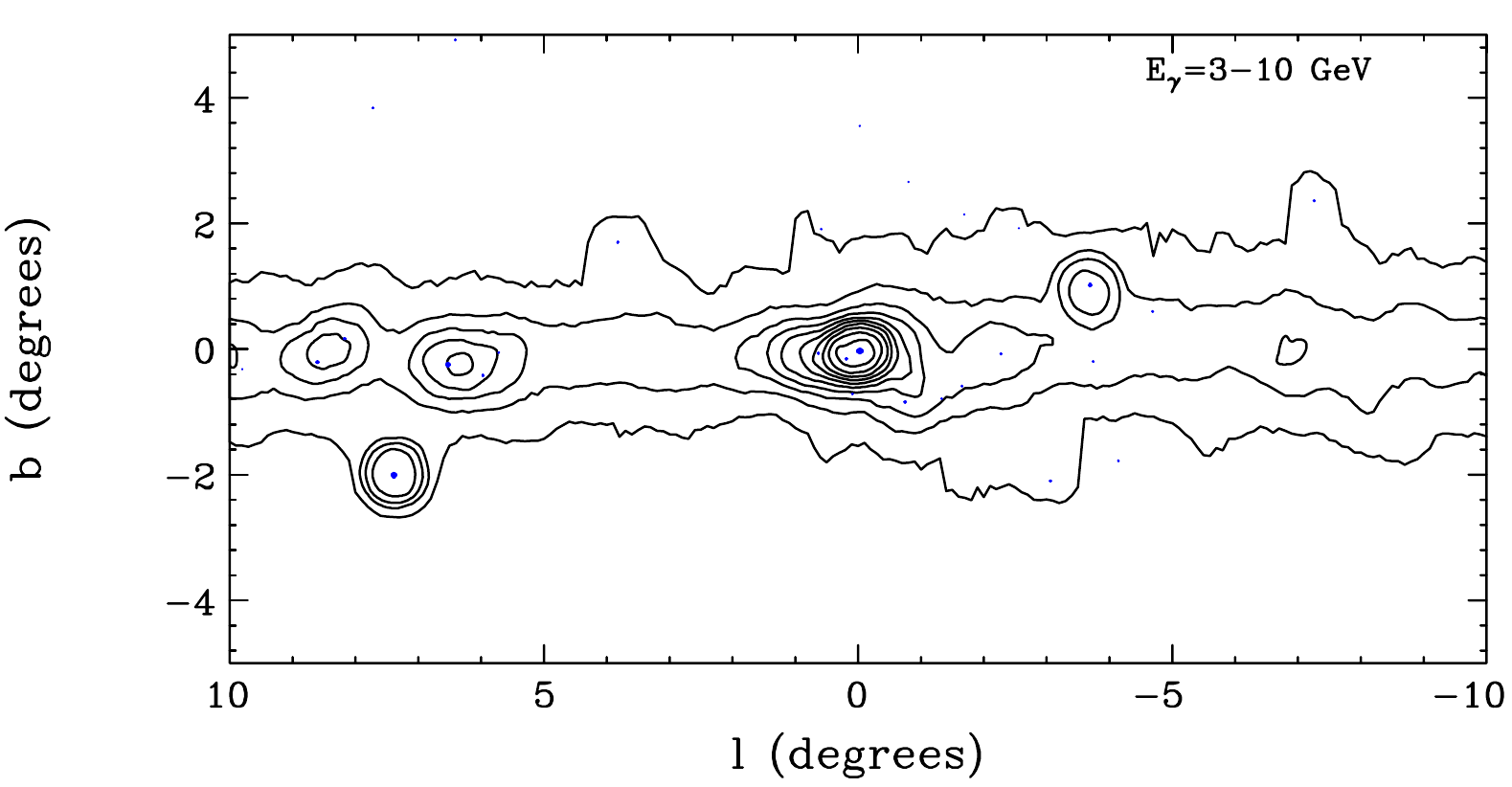}
\includegraphics[angle=0.0,width=2.2in]{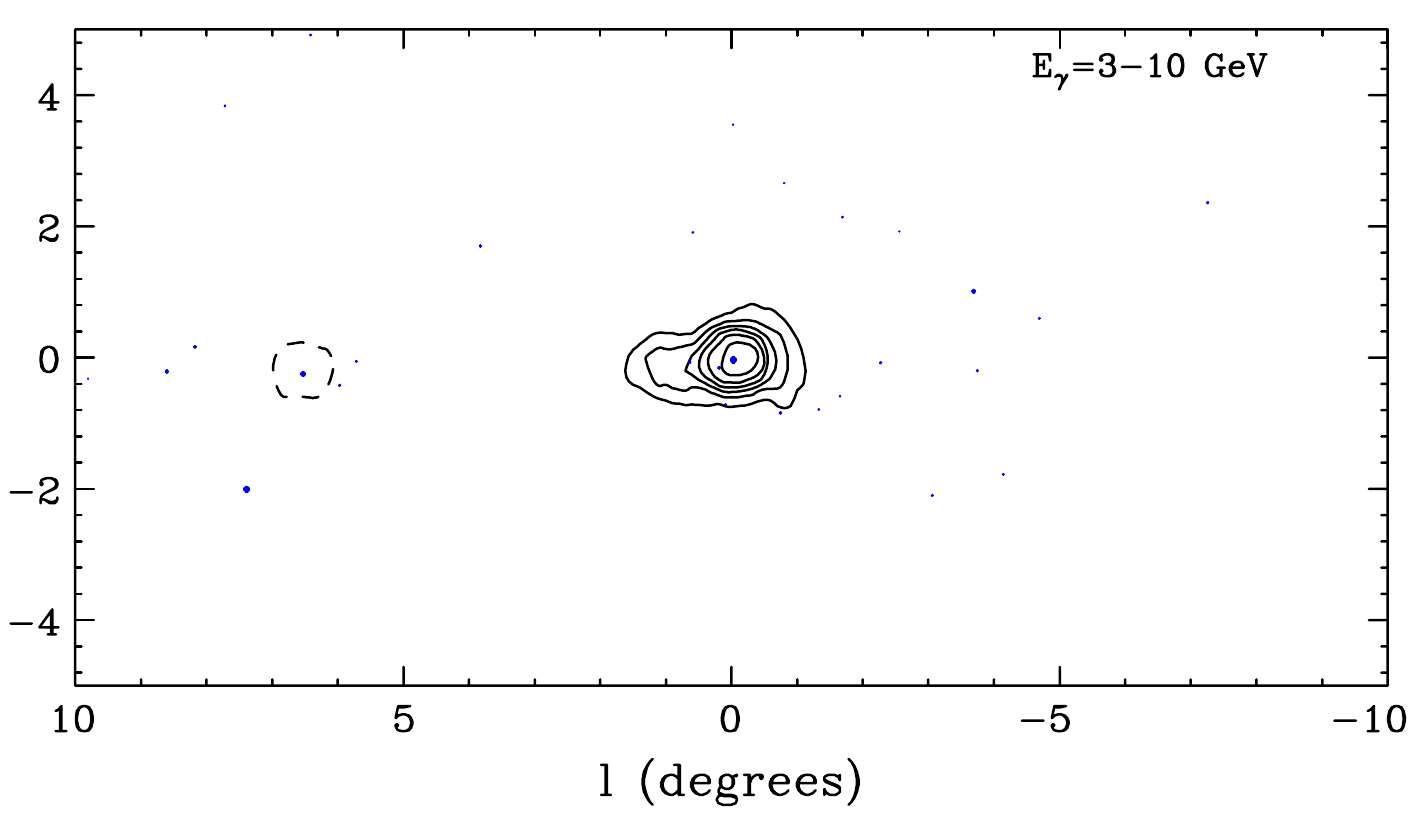}\\
\includegraphics[angle=0.0,width=2.45in]{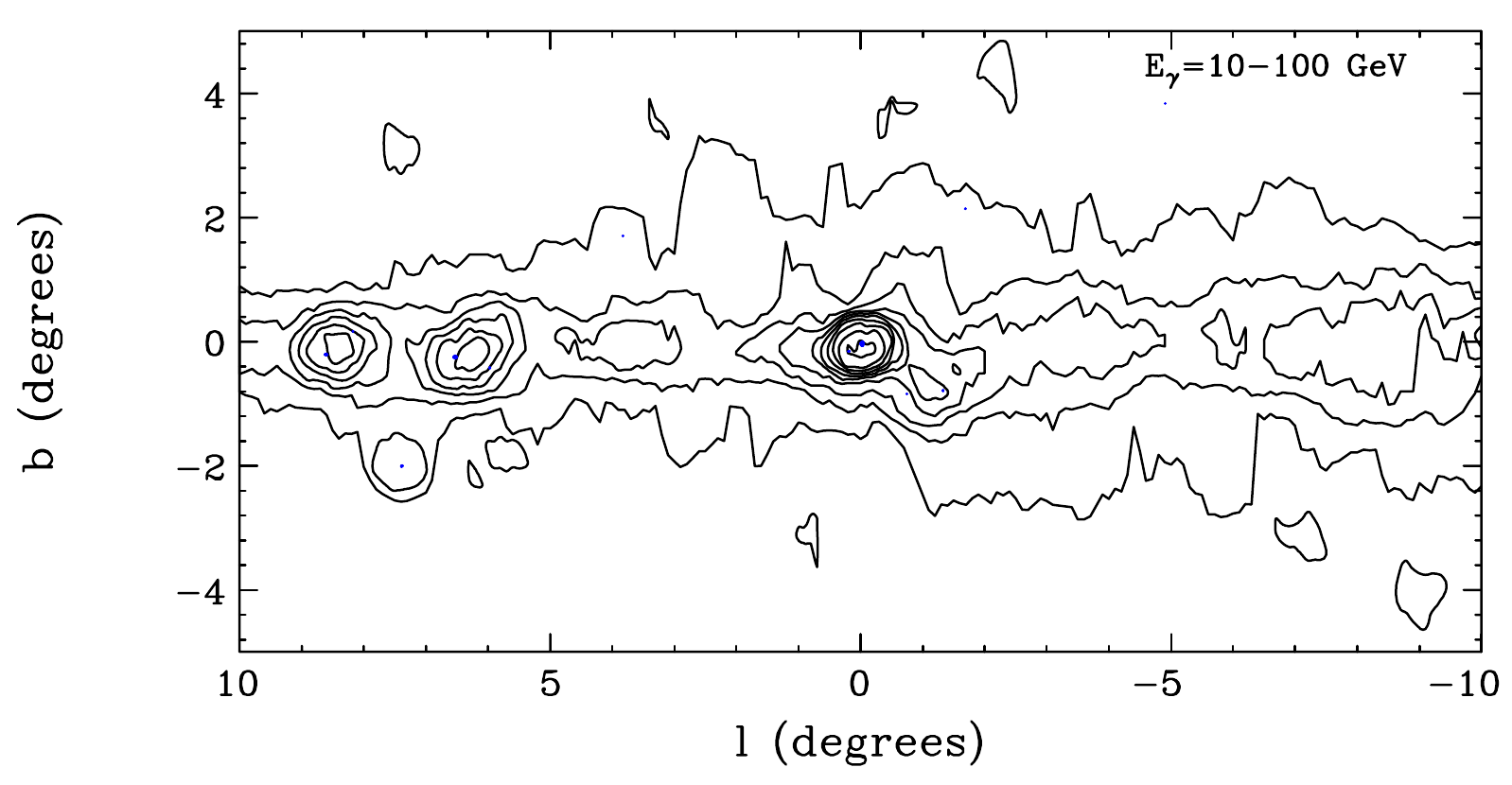}
\includegraphics[angle=0.0,width=2.2in]{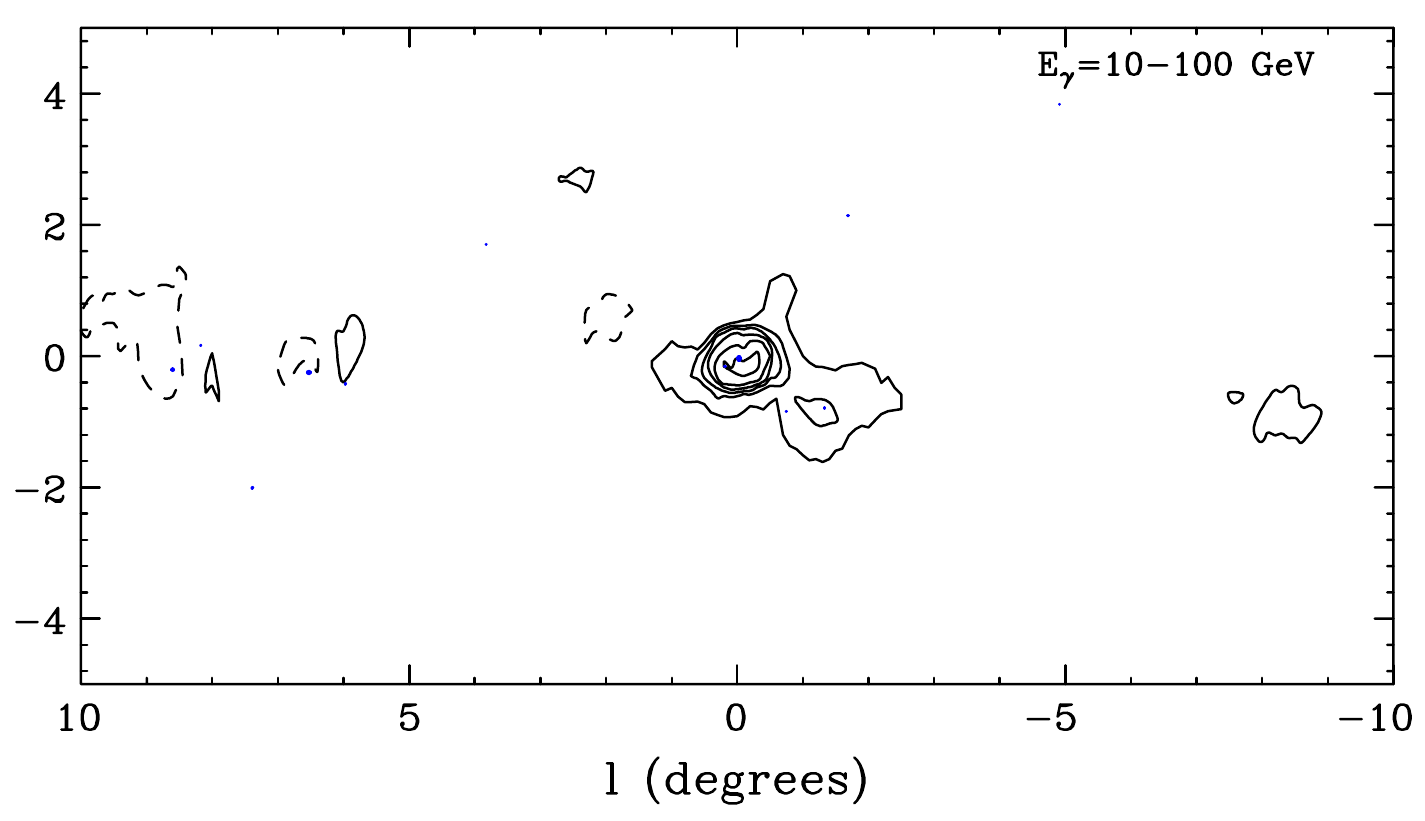}
\caption{Linearly-spaced contour maps of the gamma-ray flux from the region surrounding the Galactic Center, as observed by the Fermi Gamma-Ray Space Telescope~\cite{hooperlinden}. The left frames show the raw maps, while the right frames show the maps after subtracting known sources (not including the central source) and emission from cosmic ray interactions with gas in the Galactic Disk. This figure originally appeared in Ref.~\cite{hooperlinden}.}
\label{maps}
\end{figure*}

\begin{figure*}[t]
\centering
\includegraphics[angle=0.0,width=3.5in]{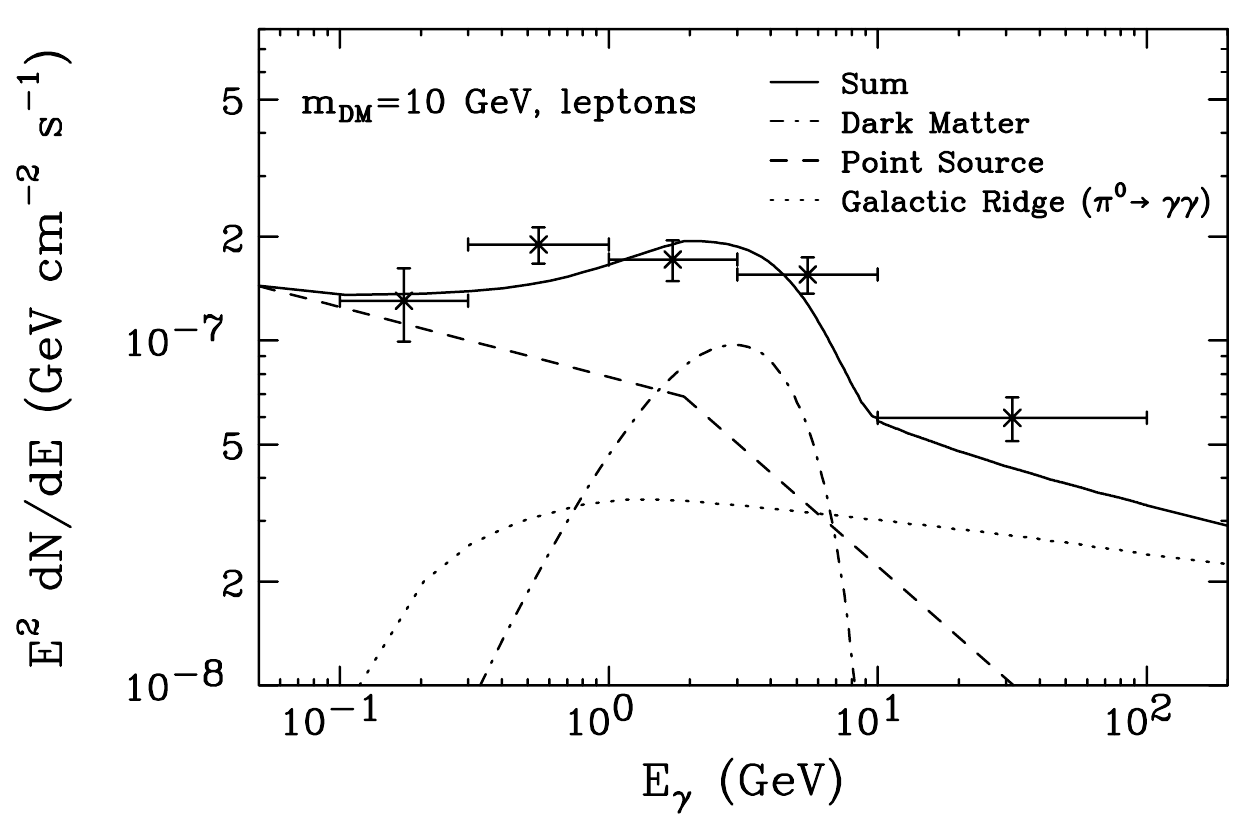}
\includegraphics[angle=0.0,width=3.5in]{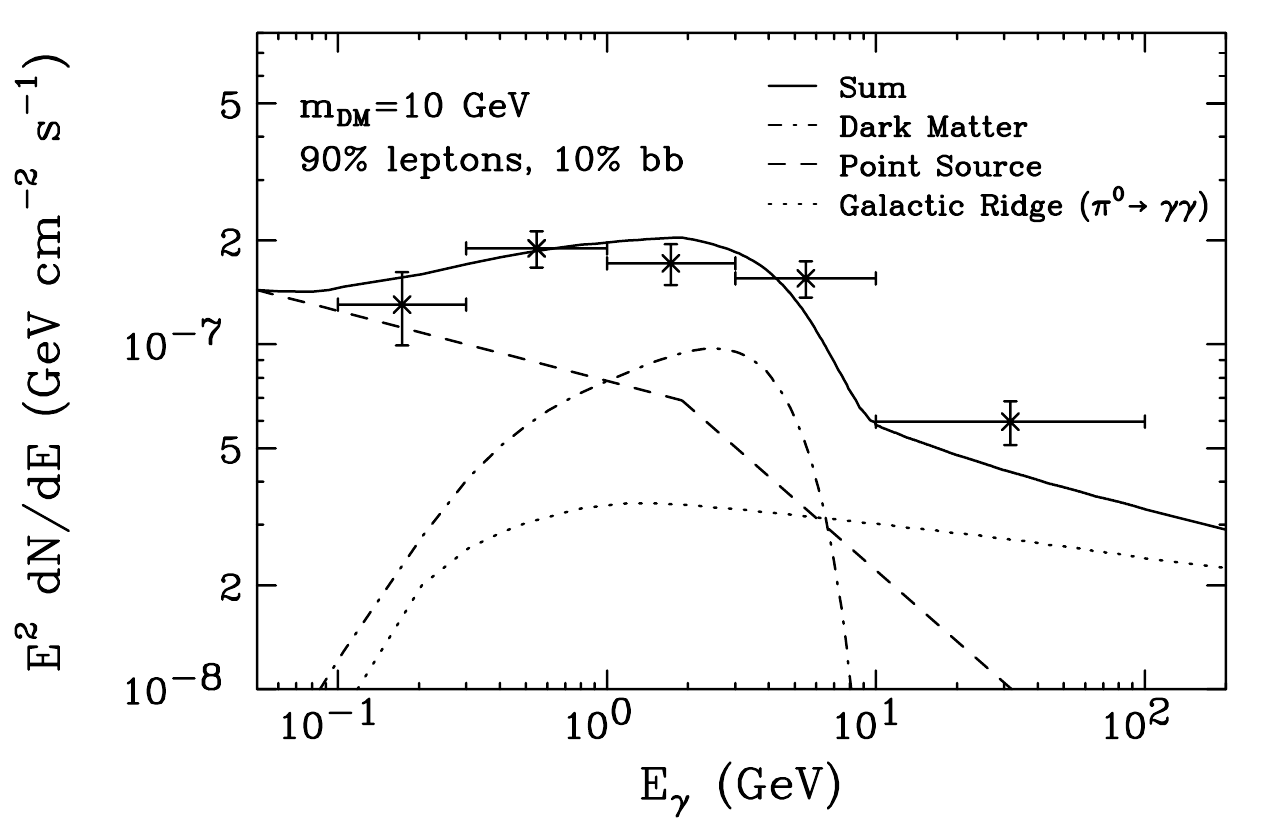}
\caption{The spectrum of residual gamma-ray emission from the inner 5 degrees surrounding the Galactic Center, after subtracting the known sources and line-of-sight gas templates. The dashed line represents the spectrum of the central, point-like emission, as found by the authors of Refs.~\cite{HG2}, \cite{Boyarsky:2010dr}, and \cite{aharonian}. Above $\sim$300 MeV, the majority of the observed emission is spatially extended, and inconsistent with originating from a point source. The dotted line shows the Galactic Ridge emission, as extrapolated from the higher energy spectrum reported by HESS~\cite{ridge}. In the left frame, I show results for a 10 GeV dark matter particle with an annihilation cross section of $\sigma v = 7\times 10^{-27}$ cm$^3$/s and which annihilates only to leptons ($e^+e^-$, $\mu^+\mu^- $ and $\tau^+ \tau^-$, $1/3$ of the time to each). In the right frame, I show the same case, but with an additional 10\% of annihilations proceeding to $b\bar{b}$. In each case, the annihilation rate is normalized to a halo profile with $\gamma=1.3$. This figure originally appeared in Ref.~\cite{hooperlinden}.}
\label{spec}
\end{figure*}

\subsection{Gamma-Rays from the Galactic Center}

Since its launch in June of 2008, the Fermi Gamma-Ray Space Telescope (FGST) has been producing the most detailed and highest resolution observations to date of the gamma-ray sky between 50 MeV and 100 GeV. In Fig.~\ref{maps}, linearly spaced contour maps of the gamma-ray emission from the region surrounding the Galactic Center are shown, derived from the first three years of Fermi data~\cite{hooperlinden}. In the left frames, raw maps are shown, smoothed at a scale of 0.5 degrees. In the right frames, two types of astrophysical backgrounds have been subtracted: known gamma-ray point sources~\cite{catalog} (shown as blue dots) and gamma-ray emission from the galactic disk. The disk model is based on the observed morphology of the disk at angles beyond $|l|=5^{\circ}$, and agree very well with observations of 21-cm surveys, which trace the density of neutral hydrogen~\cite{gas,gas2}. Note that the central bright source has not been removed, as its emission is difficult to disentangle from dark matter annihilation products originating from the inner region of a cusped halo profile. See Ref.~\cite{hooperlinden} for more details.

The gamma-ray residuals shown in the right frames of Fig.~\ref{maps} resemble in both spectrum and morphology the signal one would expect from dark matter annihilations. First of all, the angular distribution of the observed residual is spatially extended and is not consistent with that of a single point source. In Fig.~\ref{spec}, we plot the spectrum of the residual emission as shown in the right frames of Fig.~\ref{maps}. Also shown as a dashed line in Fig.~\ref{spec} is the broken power-law spectrum of point-like emission as reported by three independent groups~\cite{HG2,Boyarsky:2010dr,aharonian}. Less than half of the residual emission at energies above 300 MeV can be accounted for by a single, centrally-located point source (presumably associated with the Milky Way's supermassive black hole). Furthermore, the extended component of the emission is strongly peaked at energies between 300 MeV and 10 GeV, and drops suddenly above $\sim$10 GeV. Such a peaked spectrum is consistent with dark matter annihilation products.

To account for this spatially extended component of gamma-rays, we include in Fig.~\ref{spec} the spectrum from the annihilations of a 10 GeV dark matter particle (dot-dashed) and from a component extrapolated from HESS's observations of the Galactic Ridge (dots)~\cite{ridge}. The sum of these contributions (solid) provides a good fit to the total observed spectrum, for dark matter which annihilates mostly to leptons (the gamma-ray flux is dominated by annihilations to $\tau^+\tau^-$), possibly with a subdominant fraction proceeding to hadronic final states. To accommodate the angular extent of the observed gamma-ray signal, a dark matter distribution of approximately $\rho_{\rm DM}\propto r^{-1.25}$ to $r^{-1.4}$ is required~\cite{hooperlinden}. Interestingly, the annihilation cross section required to normalize the gamma-ray signal is not far from the value predicted for a simple thermal relic ($\sigma v = 3\times 10^{-26}$ cm$^3$/s). Adopting central values for the local dark matter density~\cite{bertonehalo}, the annihilation cross section to $\tau^+ \tau^-$ is required to be $\sigma v_{\tau\tau}\approx(1-5)\times 10^{-27}$ cm$^3$/s for a dark matter distribution with an inner slope of 1.3 to 1.4. If the dark matter also annihilates to electrons and muons at a similar rate, the total annihilation cross section falls within a factor of a few of the canonical estimate of $3\times 10^{-26}$ cm$^3$/s.\footnote{While these results are largely based on the analysis of Ref.~\cite{hooperlinden} (and its predecessors Refs.~\cite{HG1,HG2}), an independent analysis of the Fermi data in the direction Galactic Center was also presented in Ref.~\cite{Boyarsky:2010dr}. The results of Ref.~\cite{Boyarsky:2010dr} are in good agreement with those of Ref.~\cite{hooperlinden}. In particular, Ref.~\cite{Boyarsky:2010dr} find that the inclusion of a dark matter-like signal in their analysis improves the log-likelihood of their fit by 25 with the addition of only one new parameter, corresponding to a significance of approximately 5$\sigma$~\cite{Boyarsky:2010dr}. The Fermi Collaboration has also presented preliminary findings~\cite{prelim} which describe a spectrum of excess emission consistent with that found in Ref.~\cite{hooperlinden}.}

Although astrophysical origins of the gamma-ray emission observed from the Galactic Center region have been discussed~\cite{hooperlinden}, considerable challenges are faced by such interpretations. Possibilities that have been considered include emission from the central supermassive black hole~\cite{aharonian,HG2}, and from a population of unresolved point sources, such as millisecond pulsars~\cite{pulsars}. 

In the case of the supermassive black hole, direct emission from this object is not consistent with the observed morphology of the gamma-ray signal. The observed angular extent of the emission could be reconciled, however, if the gamma-rays originate from cosmic rays that have been accelerated by the black hole and then diffuse throughout the surrounding interstellar medium, producing pions through interactions with gas~\cite{aharonian,aharonianold}. The spectral shape of the spatially extended emission is very difficult to account for with gamma-rays from pion decay, however. Even for a monoenergetic spectrum of protons, the resulting spectrum of gamma-rays from pion decay does not rise rapidly enough to account for the observed gamma-ray spectrum.

A large population of unresolved gamma-ray pulsars surrounding the Galactic Center has also been proposed to account for the observed emission~\cite{HG2,pulsars,hooperlinden}. The spectra observed from among the 46 pulsars in the FGST's first pulsar catalog, however, are typically much softer than is observed from the Galactic Center~\cite{pulsarcatalog,hooperlinden}. Unless the spectra among the population of pulsars surrounding the Galactic Center is significantly different from those observed elsewhere, it does not appear to be possible to account for the observed signal with pulsars. Furthermore, it is also difficult to accommodate the very spatially concentrated morphology of the observed gamma-ray emission with pulsars. To match the observed angular distribution of this signal, the number density of pulsars would have to fall off with the distance to the Galactic Center at least as rapidly as $r^{-2.5}$. In contrast, within the innermost parsec of the Galactic Center, the stellar density has been observed to fall off only about half as rapidly, $r^{-1.25}$~\cite{Schoedel:2009mv}. Even modest pulsar kicks of $\sim 100$ km/s would allow a pulsar 10 pc from the Galactic Center to escape the region, consequently broadening the angular width of the signal. Unlike with most astrophysical sources or mechanisms, annihilating dark matter produces a flux of gamma-rays that scales with its density {\it squared}, and thus can much more easily account for the high concentration of the observed signal from the Galactic Center.

\subsection{Synchrotron Emission From The Inner Galaxy's Radio Filaments}

If dark matter annihilations produce mostly charged leptons, as implied by the Galactic Center's gamma-ray spectrum, then electrons and positrons should carry away much of the total power produced in this process. Electron and positron cosmic rays lose much of their energy to synchrotron emission, providing a potentially detectable signal for telescopes operating at radio and microwave frequencies~\cite{Fornengo:2011iq}. 

Particularly promising sources of dark matter-powered synchrotron emission are the peculiar astrophysical objects known as non-thermal radio filaments. Radio filaments are long ($\sim$40~pc) and thin ($\sim$1~pc) structures, found at distances between 10 and 200~pc from the Galactic center. The very hard spectra of highly polarized radio synchrotron emission observed from these objects~\cite{1984Natur.310..557Y} imply that they contain highly ordered poloidal magnetic fields of strength on the order of $\sim$100 $\mu$G~\cite{1986AJ.....92..818T}. These strong and highly ordered magnetic fields lead the filaments to act as magnetic mirrors, efficiently rejecting incident electrons and retaining those electrons within their volumes.

The spectrum of electrons that must be contained within the Milky Way's radio filaments in order to produce their extremely hard synchrotron emission has long been a challenge to explain astrophysically. Since the 1980s, observations of radio filaments have revealed a turnover at $\sim$10~GHz in the synchrotron spectrum, implying an electron energy spectrum that is strongly peaked (sometimes described in the radio astronomy literature as ``monoenergetic''~\cite{1988A&A...200L...9L,1992A&A...264..493L}) at an energy of approximately $\sim$10 GeV, propagating in a magnetic field on the order of 100~$\mu$G~\cite{1988A&A...200L...9L,2001RPPh...64..429M,2006ApJ...637L.101B}. The leading astrophysical mechanism proposed to explain these spectra involves magnetic reconnection zones that are formed in collisions between radio filaments and molecular clouds, leading to an electric potential capable of accelerating electrons to their required energy~\cite{1992A&A...264..493L,2004A&A...419..161L}. This scenario fails, however, to explain why so many observed radio filaments exhibit such similar spectra~\cite{2011P&SS...59..537L} (especially those without associations with molecular clouds~\cite{1999ApJ...521L..41L,2004ApJ...611..858L}). Furthermore, recent simulations find that it is unlikely that such a mechanism would be capable of accelerating electrons to energies much above 10~MeV, several orders of magnitude below that needed to explain the observed synchrotron signal~\citep{2005MNRAS.358..113L,2011arXiv1103.5924Z}.

\begin{figure}
\includegraphics[angle=0.0,width=3.3in]{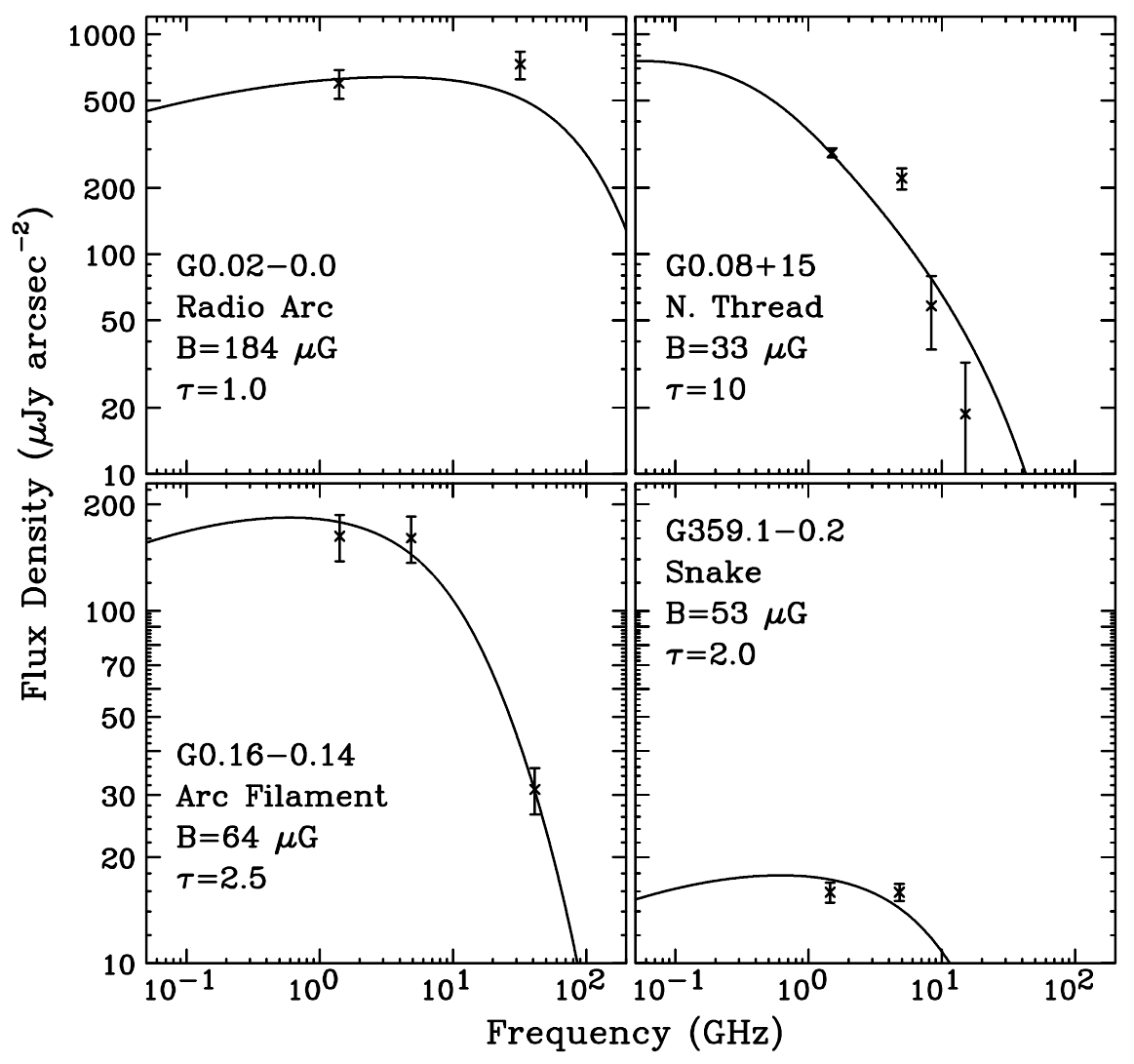}
\caption{The spectra of synchrotron emission observed from the Milky Way's non-thermal radio filaments imply that they contain a spectrum of electrons/positrons that is strongly peaked at energies near $\sim$10 GeV. Here, we compare the observed spectra of four particularly well measured radio filaments~\cite{radiospec} to that predicted from dark matter annihilations ($m_{\rm DM}=10$ GeV, annihilating equally to $e^+ e^-$, $\mu^+ \mu^-$ and $\tau^+ \tau^-$ with $\sigma v=7\times 10^{-27}$ cm$^3$~s$^{-1}$) compared to the observed intensity and spectrum of G0.2-0.0 (the Radio Arc, top left), G0.08+0.15 (Northern Thread, top right), G0.16-0.14  (Arc Filament, bottom left) and G359.1-0.2 (the Snake, bottom right). The magnetic field strengths, filamentary widths, and synchrotron energy loss times have been chosen to accommodate each filament. This figure was adapted from one originally appearing in Ref.~\cite{filaments}.}
\label{fig:4}
\end{figure}

While astrophysical mechanisms struggle to explain the strongly peaked spectrum of $\sim$10 GeV electrons present within the Milky Way's radio filaments, the annihilations of 10 GeV dark matter particles to leptons (including to $e^+ e^-$) can easily accommodate the observed spectra. In Fig.~\ref{fig:4}, the spectrum of radio emission observed from four particularly well measured filaments~\cite{radiospec} is compared to the synchrotron flux and spectrum predicted from the electrons produced through the annihilations of a 10 GeV dark matter particle. As in the previous subsection, we adopt a dark matter distribution of $\rho_{\rm DM} = 0.34 \, {\rm GeV/cm}^3 \times (r/8.5\,{\rm kpc})^{-1.3}$, a total annihilation cross section of $\sigma v=7\times 10^{-26}$ cm$^3$/s, and equal fractions of annihilations proceeding to $e^+ e^-$,  $\mu^+ \mu^-$ and $\tau^+ \tau^-$. For each filament, we have adopted values of the magnetic field ($B$) and the ratio of timescales for diffusion and synchrotron losses ($\tau$) which best accommodate the observed spectra. In each case, we find it possible to provide a good fit to the filament's spectral shape. As the overall normalization of the emission from each filament is proportional to its overall volume, there is some uncertainty in the overall intensity predicted from each such object. For each filament, we have treated their geometry as cylindrical and have adopted a width which provides the necessary normalization. In each case, the width we have used is within a factor of two of that reported in the observational literature. This modest discrepancy could very plausibly be accounted for by factors such as differences between observed peak luminosities (as are typically reported) and average luminosities (being calculated here).

A particularly powerful test of the dark matter interpretation for the emission of radio filaments is the strong dependence of the luminosity with distance to the Galactic Center that is predicted. For a dark matter distribution of the form $\rho_{\rm DM} \propto r^{-1.3}$ (as motivated by the morphology of the Galactic Center gamma-ray signal), a filament at a distance of 10 pc from the Galactic Center should be nearly 400 times brighter than an identical filament located at 100 pc from the center. Although variations in each filament's geometry, magnetic field, and other properties will lead to a degree of filament-to-filament scatter in their overall brightness, if dark matter annihilations are in fact powering the radio emission of these objects, a strong correlation with galactocentric distance should be evident.

\begin{figure}
\includegraphics[angle=0.0,width=3.3in]{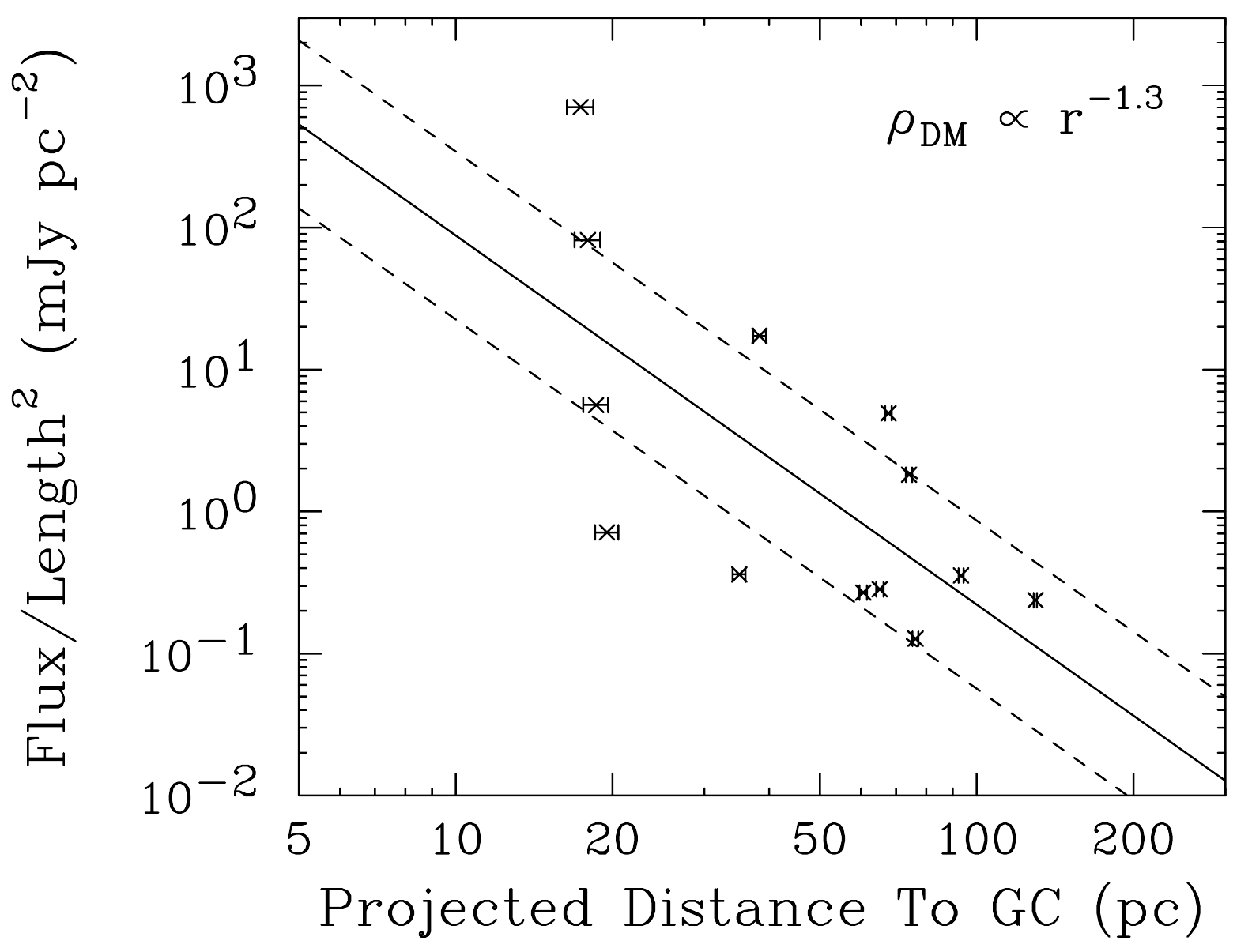}
\caption{Flux per length squared (in milli-Janskys per square parsec) for 13 radio filaments at 1.4 GHz~\cite{1pt4}, as a function of their projected distance from the Galactic Center. There is a clear trend that those filaments closer to the Galactic Center are brighter than those farther away. The solid line denotes the slope of this correlation predicted for a dark matter distribution of the form $\rho_{\rm DM} \propto r^{-1.3}$. The dashed lines shown in Fig.~\ref{fig:5} were chosen such that they bracket 68\% of the projected directions from the Galactic Center, and can be thought of as an uncertainty intrinsic to this comparison. This figure was adapted from one originally appearing in Ref.~\cite{filaments}.}
\label{fig:5}
\end{figure}

In Fig.~\ref{fig:5}, we compare the flux divided by the square of the filament's length to the projected distance of the filament to the Galactic Center, for 13 filaments measured at 1.4 GHz~\cite{filaments,1pt4} (the length squared factor is included to account for the greater volume of longer filaments and the longer length of time that longer filaments will retain electrons, valid in the regime in which $\tau$ is of order unity). We find a very significant correlation, consistent with that predicted for annihilating dark matter distributed as $\rho_{\rm DM} \propto r^{-1.3}$. As we are plotting the projected distance rather than the actual (but unknown) distance to the Galactic Center, we expect a sizable amount of scatter to appear. The dashed lines shown in Fig.~\ref{fig:5} were chosen such that they bracket 68\% of the projected directions from the Galactic Center, and can be thought of as an uncertainty intrinsic to this comparison. Note that observations of radio filaments at frequencies below that expected to be dominated by dark matter (330 MHz, in particular) do not exhibit this correlation~\cite{filaments}.

\subsection{Synchrotron Emission From The Inner Galaxy: ``The WMAP Haze''}

If annihilating dark matter particles are responsible for the gamma-ray and radio signals described in the previous two subsections, then a sizable flux of energetic electrons and positrons will be injected into the Inner Galaxy. And whereas any electrons and positrons produced within the volumes of radio filaments will have their energy rapidly converted into the hard and polarized synchrotron emission observed from those filaments, dark matter annihilations taking place elsewhere in the inner Milky Way will produce a more diffuse synchrotron signal through interactions with the Galactic Magnetic Field.

In addition to cosmic microwave background photons, the WMAP (Wilkinson Microwave Anisotropy Probe) experiment has provided the best measurements to date of a number of standard emission mechanisms known to take place in the interstellar medium, including emission from thermal dust, spinning dust, ionized gas, and synchrotron~\cite{spergel}. GeV-scale cosmic ray electrons in the presence of 10-50$\mu$G magnetic fields produce synchrotron emission that peaks at GHz frequencies, and within the frequency range studied by WMAP and other CMB experiments. WMAP's observations have revealed an excess of microwave emission in the inner $20^{\circ}$ around the center of the Milky Way, distributed with approximate radial symmetry, and uncorrelated with all other known foregrounds~\cite{haze1,doblerfink}. This anomalous emission, known as the ``WMAP Haze'', is generally interpreted as hard synchrotron emission from a population of energetic cosmic ray electrons/positrons present in the inner kiloparsecs of the Milky Way. Due to the morphology and overall power of the WMAP Haze, it has been proposed that this signal could be synchrotron emission from electrons and positrons produced through dark matter annihilations~\cite{darkhaze,darkhaze1,timhaze}.\footnote{More recently, a diffuse flux of gamma-rays has been identified at high latitudes in the Fermi data, likely resulting from the Inverse Compton scattering of $\sim$TeV electrons/positrons~\cite{Su:2010qj} (or possibly the scattering of cosmic ray hadrons with gas~\cite{Crocker:2010dg}). While it is possible that this emission (which goes by names such as the Fermi Haze, the Fermi Bubbles, and the Fermi Lobes) is in some way connected to the WMAP Haze, it is also possible that these signals result from two separate populations of cosmic rays, with considerably differing energies and which are evident in quite different parts of the sky.}

To calculate the synchrotron signal predicted from the annihilations of 10 GeV dark matter particles, one must model the propagation of the electron and positron annihilation products through the inner galaxy. We do this using the cosmic ray propagation code $Galprop$~\cite{galprop}, adopting conventional values for the diffusion coefficient ($3.5\times 10^{28}$ cm$^2$/s) and Galactic Magnetic Field ($B = 22 \, \mu{\rm G} \,\, e^{-r/5.0\,{\rm kpc}} \, e^{-|z|/1.8\,{\rm kpc}}$, where $r$ and $z$ represent the distance from the Galactic Center along and perpendicular to the the Galactic Plane).

\begin{figure}\mbox{\includegraphics[width=0.5\textwidth,clip]{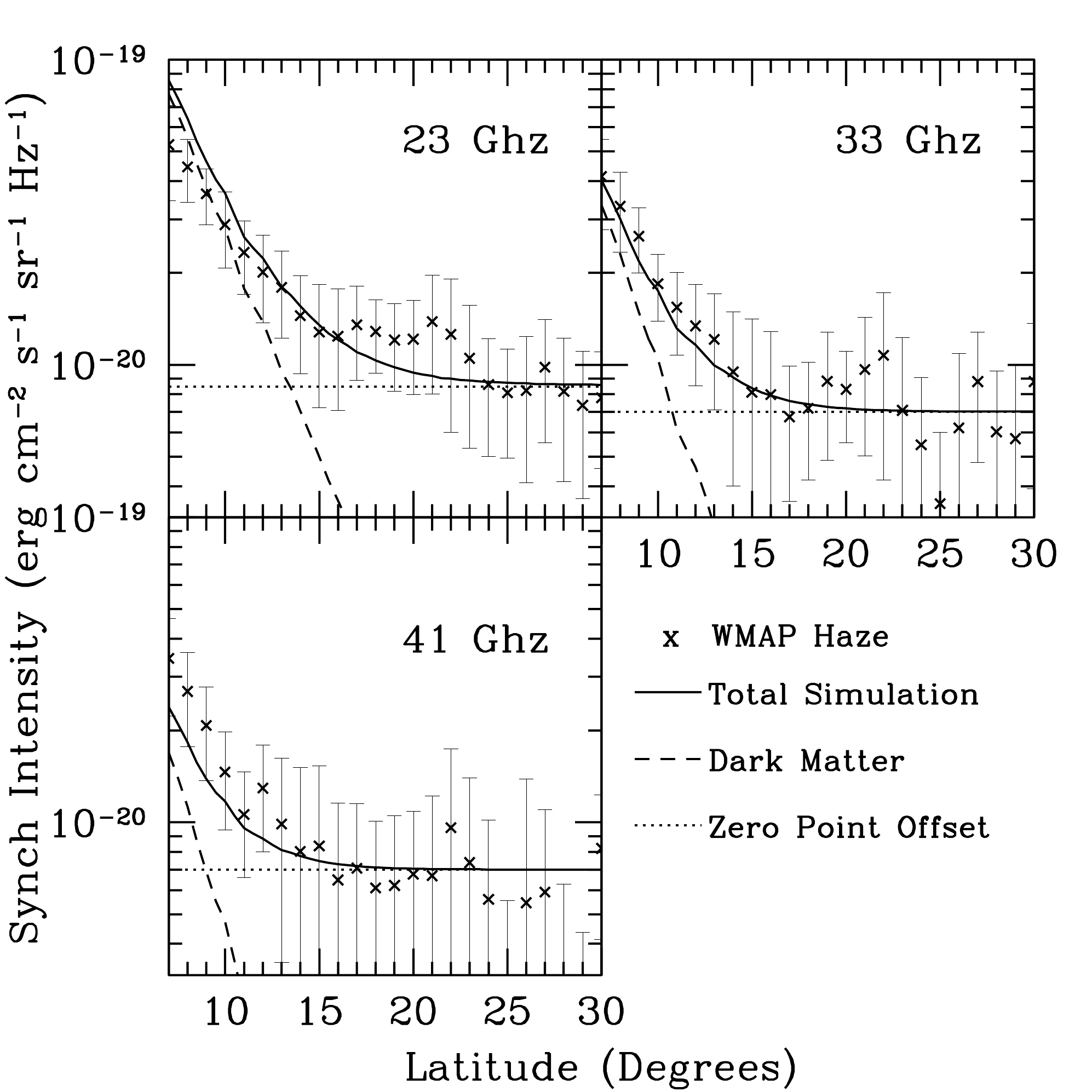}}\caption{Synchrotron emission from dark matter annihilations as a function of latitude below the Galactic Center for 10 GeV dark matter particles annihilating equally to $e^+ e^-$, $\mu^+ \mu^-$, and $\tau^+ \tau^-$, distributed as $\rho_{\rm DM} = 0.35 \, {\rm GeV/cm}^3 \times (r/8.5\,{\rm kpc})^{-1.33}$, and with a total cross section of $\sigma$v~=~7~x~10$^{-27}$ cm$^3$/s. The magnetic field model used is given by $B(r,z) = 22 \, \mu{\rm G} \,\, e^{-r/5.0\,{\rm kpc}} \, e^{-|z|/1.8\,{\rm kpc}}$. This figure was adapted from one originally appearing in Ref.~\cite{timhaze}.}
\label{democratic}
\end{figure}

In Fig.~\ref{democratic}, we compare the synchrotron haze predicted from 10 GeV dark matter particles to that observed by WMAP. Here, we have used the same dark matter model as in the previous two subsection (with the exception of a slightly different distribution, $\rho_{\rm DM}\propto r^{-\gamma}, \gamma=1.33$ rather than $\gamma=1.3$, which should be of little consequence). We find quite good agreement with the observed features of the WMAP Haze. These fits to the WMAP Haze were obtained with relatively little freedom in the astrophysical or dark matter parameters. In particular, the mass, annihilation cross section, and halo profile are each tightly constrained by the observed features of the Galactic Center gamma-ray signal. Although the choice of the magnetic field model allowed us to adjust the morphology and spectrum of the of the synchrotron emission to a limited degree, we had little ability to significantly adjust the overall synchrotron intensity. If the gamma-rays from the Galactic Center as observed by Fermi are interpreted as dark matter annihilation products, we are forced to expect a corresponding synchrotron signal from the Inner Galaxy very much like that observed by WMAP.

Dark matter particles annihilating in galaxies other than the Milky Way will produce annihilation products which contribute to the diffuse isotropic radio background. Interestingly, data from ARCADE 2 (Absolute Radiometer for Cosmology, Astrophysics and Diffuse Emission), and a number of low frequency radio surveys have revealed a sizable flux of isotropic power at radio frequencies ($\lsim 3$ GHz), brighter than a factor of 5-6 than that expected based on extrapolations of of the luminosity functions of known radio sources. This emission also exhibits a harder spectrum than is observed from resolved sources such as radio galaxies~\cite{arcade}. In Ref.~\cite{arcadedarkmatter} it was suggested that dark matter annihilations may account for this excess. In particular, they point out that 10 GeV dark matter particles annihilating to leptons can provide a good fit to the observed radio background, without relying on large boost factors~\cite{arcadedarkmatter,pc}.

\subsection{Indirect Evidence Summary and Constraints}

Over the past several pages, I have summarized three independent astrophysical observations which can be explained by the annihilations of a 10 GeV dark matter particle (four if you include the excess power in the diffuse radio background). In this subsection, I will briefly discuss what these observations (if interpreted as dark matter annihilation products) tell us about the dark matter particle and its distribution, and compare this to various constraints that can be placed from other observations.

Beginning with the dark matter distribution, the angular distribution or morphology of the gamma-ray signal observed from the Galactic Center requires a dark matter distribution of $\rho_{\rm DM} \propto r^{-\gamma}$, with $\gamma\approx1.25$ to 1.40~\cite{hooperlinden}. This is consistent with the correlation observed between the geometrically corrected flux from the Milky Way's radio filaments and their projected distance to the Galactic Center (see Fig.~\ref{fig:5}). More specifically, the best-fit linear regression for this correlation favors a slope of $\rho_{\rm DM} \propto r^{-1.27}$~\cite{filaments}. The observed morphology of the WMAP Haze is also consistent with this dark matter distribution (see Fig.~\ref{democratic}), although this is somewhat degenerate with the choice of the distribution of the Galactic Magnetic Field in the inner kiloparsecs of the Milky Way. If we assume that this profile slope persists out to the solar neighborhood, we can combine these results with rotation curve and microlensing measurements~\cite{bertonehalo} to arrive at $\rho_{\rm DM} \approx 0.34 \, {\rm GeV/cm}^3 \times (r/8.5\,{\rm kpc})^{-1.3}$, although with significant uncertainties in the overall normalization. One should keep in mind that the slope of the profile could be different that the value favored by these observations at distances beyond a few kiloparsecs from the Galactic Center.

To accommodate the gamma-ray spectrum observed from the Galactic Center, we require either a dark matter particle with a mass in the range of 7-12 GeV which annihilates primarily to leptons (including to $\tau^+ \tau^-$) or a mass in the range of 25-45 GeV and which annihilates to hadronic final states (such as to $b\bar{b}$). To produce the synchrotron spectrum observed from the Milky Way's radio filaments, however, we must require the production of $\sim$10 GeV electrons and thus choose to focus on the lower of these mass ranges. If the dark matter particles annihilate roughly equally to $e^+ e^-$ and $\tau^+ \tau^-$ (along with any annihilations to muons), the spectra and relative normalizations of the Galactic Center gamma-ray, radio filaments, and WMAP Haze signals can all be accommodated simultaneously. A modest fraction ($\lsim$ 20\%) of annihilations could also proceed to quarks, although this is not required. Using the previously mentioned dark matter distribution, and assuming equal fractions of annihilations proceed to $e^+ e^-$, $\mu^+ \mu^-$, and $\tau^+ \tau^-$, the normalization of these signals requires an annihilation cross section of $\sigma v \approx 7\times 10^{-27}$ cm$^3$/s. Again, uncertainties in the dark matter distribution make this quantity uncertain at the level of a factor of a few. 

We can now compare the dark matter properties inferred from these signals to the various constraints that can be derived from other observations.

\begin{figure*}[!t]
\centering
{\includegraphics[angle=0.0,width=2.4in]{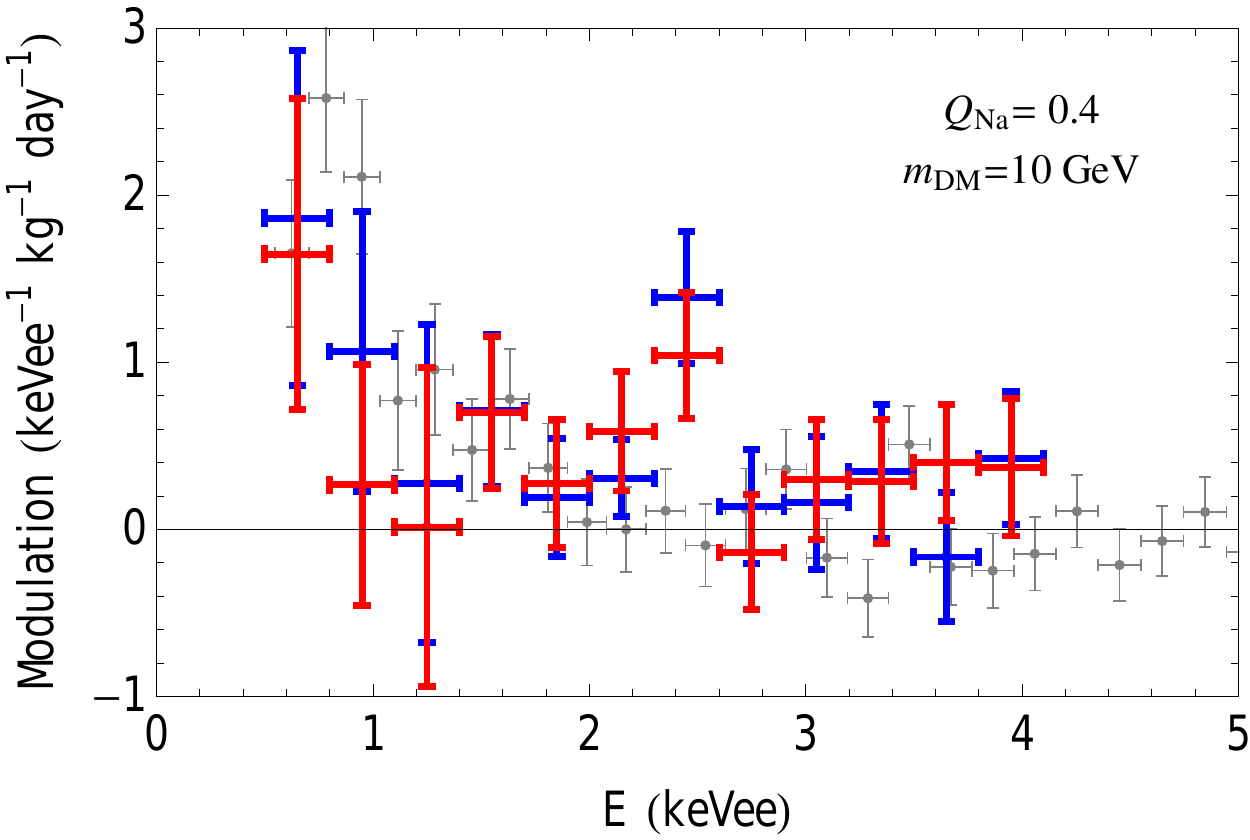}}
{\includegraphics[angle=0.0,width=2.25in]{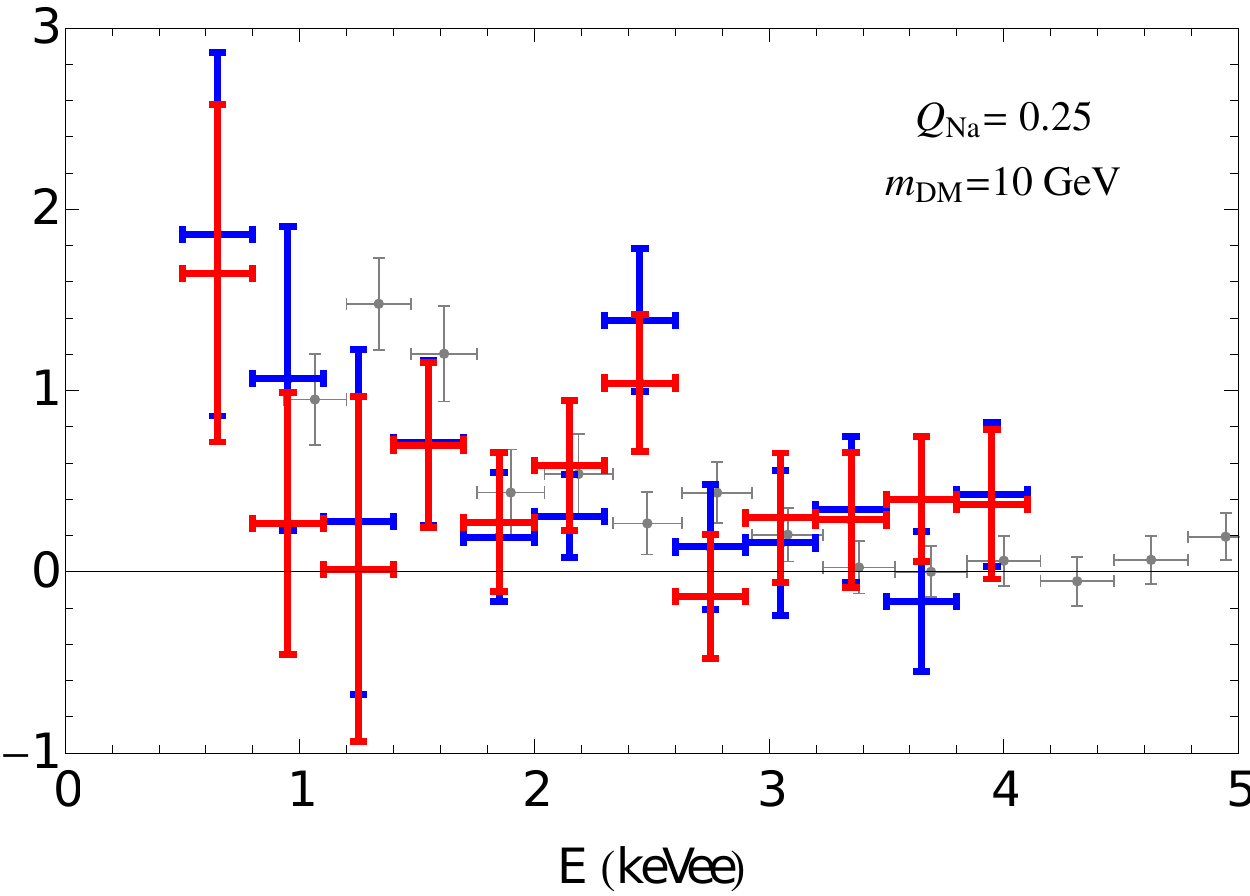}}
{\includegraphics[angle=0.0,width=2.25in]{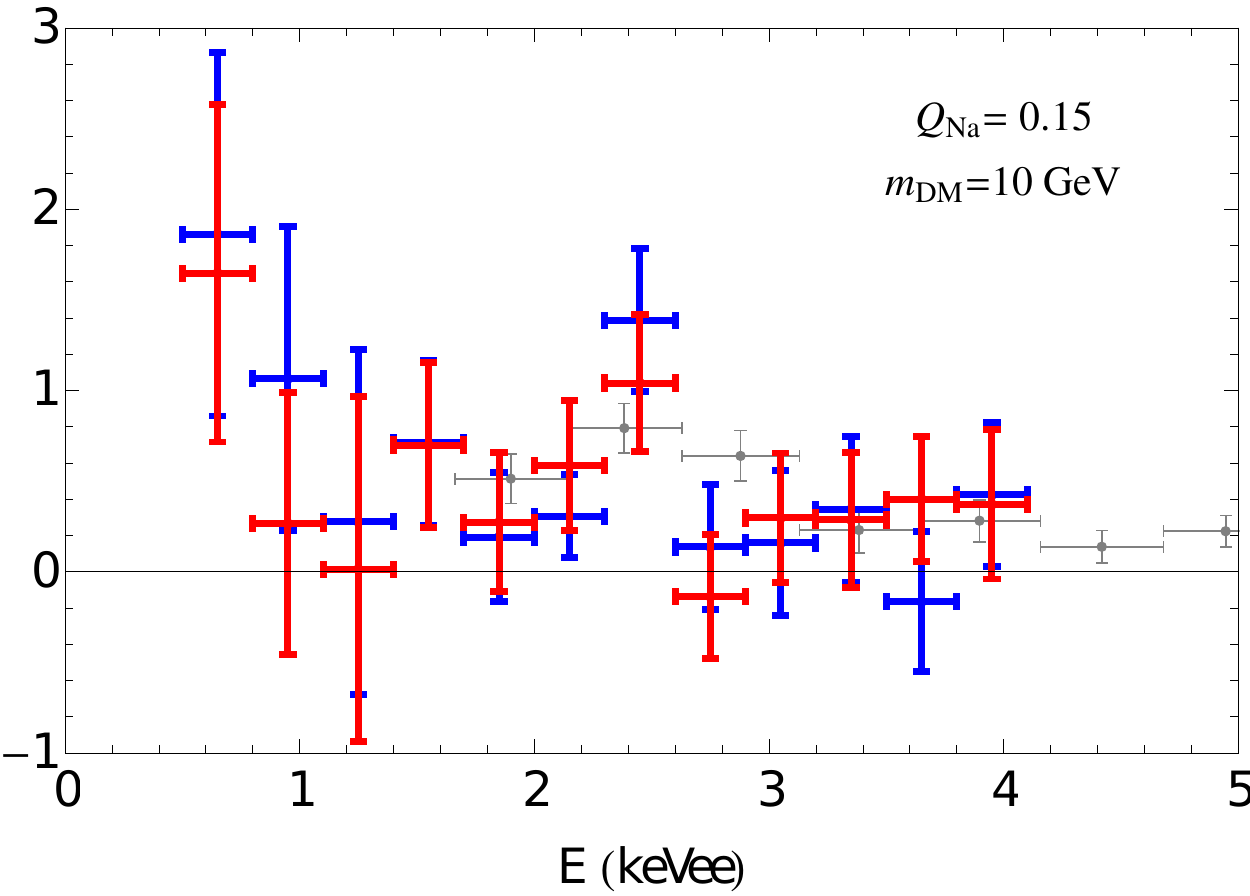}}
\caption{A comparison between the modulation amplitude spectrum observed by DAMA/LIBRA and CoGeNT, independent of the dark matter's velocity distribution, following the approach of Ref.~\cite{Fox:2010bz}. The comparison is done for a dark matter mass of 10 GeV, and for three choices of the low-energy sodium quenching factor. The blue (red) error bars denote the CoGeNT modulation amplitude assuming a phase that peaks on April 18th (May 26th). The grey error bars denote the DAMA/LIBRA modulation spectrum. In normalizing the results, we have assumed the dark matter's elastic scattering cross section to scale with the square of the target's atomic number, $A^2$. This figure originally appeared in Ref.~\cite{arXiv:1110.5338}.}
\label{fig:damaAtCogent}
\end{figure*}

\begin{itemize}
\item{{\bf Gamma-rays from dwarf spheroidal galaxies:} The Fermi collaboration's combined analysis of 10 dwarf spheroidals excludes 10 GeV dark matter particles annihilating to $\tau^+ \tau^-$ with an annihilation cross section greater than $\sigma v_{\tau\tau} \approx 1.4\times 10^{-26}$ cm$^3$/s (or $\sigma v \approx 4.2 \times 10^{-26}$ cm$^3$/s for equal fractions of annihilations to electrons, muons and taus)~\cite{dwarf}. This is a factor of approximately six larger than that required to normalize the signals discussed in this section. Other gamma-ray constraints, such as those from galaxy clusters~\cite{clusters} or from the diffuse gamma-ray background~\cite{cosmo} are also not yet sensitive to the annihilation cross sections required in this scenario.}

\item{{\bf Effects on the cosmic microwave background:} Dark matter annihilation products can heat and ionize the photon-baryon plasma at $z\sim1000$, distorting the CMB. For a 10 GeV dark matter particle annihilating equally to $e^+ e^-$, $\mu^+ \mu^-$, and $\tau^+ \tau^-$, however, WMAP only requires $\sigma v \lsim 8 \times 10^{-26}$ cm$^3$/s~\cite{Finkbeiner:2011dx}, an order of magnitude above the value required here. This effect could potentially be within the reach of Planck, however.}

\item{{\bf Neutrinos from the Sun:} Dark matter particles can elastically scatter with nuclei in the Sun, leading to their gravitational capture and subsequent annihilation. Electrons and muons produced in such annihilations quickly lose their energy to the Solar medium and produce no observable effects. Annihilations to taus, on the other hand, produce neutrinos which can be observed by Super-Kamiokande. For a 10 GeV dark matter particle annihilating 1/3 of the time to $\tau^+ \tau^-$, Super-Kamiokande data can be used to constrain the dark matter's spin-independent elastic scattering cross section with protons to be less than $\sigma \approx 4 \times 10^{-41}$ cm$^2$~\cite{Kappl:2011kz}, which is larger than the value required to accommodate CoGeNT, CRESST-II and DAMA/LIBRA (see Sec.~\ref{direct}).}

\item{{\bf Cosmic ray antimatter:} Measurements of the antiproton and positron components of the cosmic ray spectrum can be used to place constraints on the dark matter annihilation rate in the Galactic Halo. As the dark matter particle being considered here annihilates to hadronic final states no more than $\sim$10-20\% of the time, current cosmic ray antiproton constraints from PAMELA and BESS Polar-II are about an order of magnitude weaker than would be needed to test this scenario~\cite{Kappl:2011jw}. Dark matter annihilating directly to $e^+ e^-$ can lead to a distinctive edge-like feature in the cosmic ray positron fraction (and potentially in the cosmic ray electron+positron spectrum) at an energy equal to the mass of the dark matter particle~\cite{Baltz:2004ie}. For reasonable estimates of the local radiation and magnetic field densities ($\rho_{{\rm Rad}+B} \sim 2$ eV/cm$^3$), the model being considered here will lead to a cosmic ray positron flux of $E_{e^+}^3 dN_{e^+}/dE_{e^+} \sim 1.6$ GeV$^2$ m$^{-2}$ s$^{-1}$ sr$^{-1}$ at 10 GeV, corresponding to a sub-percent feature in the positron fraction, which would be difficult to identify with PAMELA~\cite{Adriani:2008zr}. It is conceivable, however, that AMS-02 may be able to resolve such a feature. Note that the anomalous positron fraction reported by PAMELA~\cite{arXiv:0810.4995}  and confirmed by Fermi~\cite{FermiLAT:2011ab}  is observed at much higher energies that is being considered here.}

\end{itemize}

\section{Evidence From Direct Detection}
\label{direct}

\subsection{Annual Modulation: DAMA/LIBRA and CoGeNT}

If a population of events observed in a detector result from the elastic scattering of dark matter particles, then the Earth's motion around the Sun will induce a degree of seasonal variation in the rate of those events~\cite{modulation}. This signature provides a way of discriminating a dark matter signal from various backgrounds which either do not undergo annual modulation, or that modulate with a different phase or period than is predicted for dark matter. 


For over a decade, the DAMA collaboration has been reporting an annual modulation in their event rate, consistent in phase and period with that expected from dark matter~\cite{INFN-AE-00-01,damanew}. More recently, the larger DAMA/LIBRA detectors (consisting of 242.5 kg of high purity NaI(Tl) crystals) have observed annual modulation with a statistical significance of 8.9$\sigma$~\cite{damanew}. The variation of DAMA/LIBRA's rate is consistent with a sinusoid peaking at May 16$\pm$7 days at energies between 2 and 4 keV, May 22$\pm$7 days between 2 and 5 keV, and May 26$\pm$7 days between 2 and 6 keV, within the range predicted for dark matter based on the results of numerical simulations~\cite{Kuhlen:2009vh}.

Since the time of DAMA's original claim, null results from CDMS and other experiments have ruled out much of the dark matter parameter space which could potentially account for their signal. Exceptions include models in which dark matter scatters inelastically with nuclei~\cite{inelastic}, and models in which the dark matter is relatively light ($m_{\rm DM} \sim 5-20$ GeV)~\cite{lightorstream,Bottino:2003cz}.

In May of 2011, the CoGeNT collaboration reported evidence of an annual modulation in their event rate, although with a modest statistical significance of 2.8$\sigma$~\cite{Aalseth:2011wp} (they have also reported an overall excess of low energy events~\cite{Aalseth:2010vx,JuanTAUP}, which we will return to later in this article). Despite the limited statistical significance of this signal, it shares many of the features possessed by DAMA's modulation. The peak of CoGeNT's phase is April 18$\pm$16 days, which is slightly earlier (at the 1.6$\sigma$ level) than that favored by DAMA/LIBRA. A common phase that peaks in early May would be consistent with both experiments~\cite{Hooper:2011hd} and with expectations for dark matter. 

In comparing the spectrum and amplitude of DAMA and CoGeNT's annual modulation signals, it is possible to plot the results in such a way that does not depend on the velocity distribution of the dark matter particles in the local halo~\cite{Fox:2010bz,Fox:2010bu}. In Fig.~\ref{fig:damaAtCogent}, we compare the modulation spectrum as observed by CoGeNT (for two different choice of phase) to the equivalent spectrum from DAMA/LIBRA. As the quenching factor for low-energy sodium recoils is quite uncertain~\cite{damaquenching,consistent,JuanTAUP}, results are shown for three values of this quantity ($Q_{\rm Na}=0.40$, 0.25, and 0.15). To generate this comparison, we adopt $m_{\rm DM}=10$ GeV, and an elastic scattering cross section which scales as $A^2$, where $A$ is the atomic mass of the target nucleus (valid for spin-independent scattering with equal couplings to protons and neutrons). From this figure, it is evident that the overall spectrum and normalization of the modulation amplitudes reported by CoGeNT and DAMA/LIBRA are in good agreement, although with sizable uncertainties associated with the sodium quenching factor. If the modulation reported by DAMA/LIBRA is the product of dark matter spin-independent elastic scattering, then one should expect CoGeNT to observe a modulation with broad features very much like that they report, and vice versa. As CoGeNT (and its planned extension CoGeNT-4) accumulates more data, it will become increasingly possible to make detailed comparisons between the modulation spectra observed by these two experiments~\cite{kelso}.


Given the similarities  between the modulations observed by DAMA/LIBRA and CoGeNT, it appears unlikely that these signals are detector effects, or the result of the experiments' local environments (DAMA/LIBRA and CoGeNT make use of different detector materials and are located on different continents). If their signals do not arise from dark matter scattering, they are most likely associated with a common modulating background. Potential backgrounds which are known to exhibit seasonal variation consist of those associated with the underground muon flux and resulting from radon decays~\cite{modbg}. Neither of these possibilities appear to possess the characteristics required to produce the signals observed by DAMA/LIBRA and CoGeNT, however. Although the underground muon flux is known to modulate as a result of seasonal variations of the temperature and density of the upper atmosphere, the phase of this modulation has been measured to peak on July 5$\pm$15 at Gran Sasso~\cite{LVD} and July $(7-9)\pm3$ at Soudan~\cite{minos,minosslides}, each of which are inconsistent with the phases reported by DAMA/LIBRA and CoGeNT~\cite{Chang:2011eb}. Furthermore, simple estimates of the muon-induced neutron flux lead to a rate that is $\sim$$10^{-2}$ to $10^{-3}$ times smaller than required to generate DAMA/LIBRA's modulation. The phase of radon-induced backgrounds has also been measured to peak considerably later than DAMA/LIBRA's signal, in August/September~\cite{minosslides,radon}. Further challenging these interpretations is the fact that DAMA/LIBRA's multiple hit events (which are attributed to neutrons) do not show evidence of modulation. To date, no background has been identified with characteristics (phase, spectrum, and rate) compatible with the signals observed by DAMA/LIBRA and CoGeNT (see also, Ref.~\cite{Bernabei:2009du}).

\subsection{Excess Low-Energy Events: CoGeNT and CRESST-II}

In addition to their detection of an annual modulation in their event rate, the CoGeNT collaboration has also reported the observation of an overall excess of low-energy events~\cite{Aalseth:2010vx}. While some of these events are thought to be unidentified surface events, this background does not appear to be sufficient to account for the observed low-energy rate~\cite{JuanTAUP}. Even more recently, the CRESST-II collaboration has reported a similar excess of low-energy nuclear recoil candidate events~\cite{Angloher:2011uu}. In their analysis, they identified 67 low-energy nuclear recoil candidate events, which is at least 30\% more than can be accounted for with known backgrounds. The CRESST-II collaboration has assessed the statistical significance of their excess to be greater than 4$\sigma$.

\begin{figure}[!t]
\centering
{\includegraphics[angle=0.0,width=3.3in]{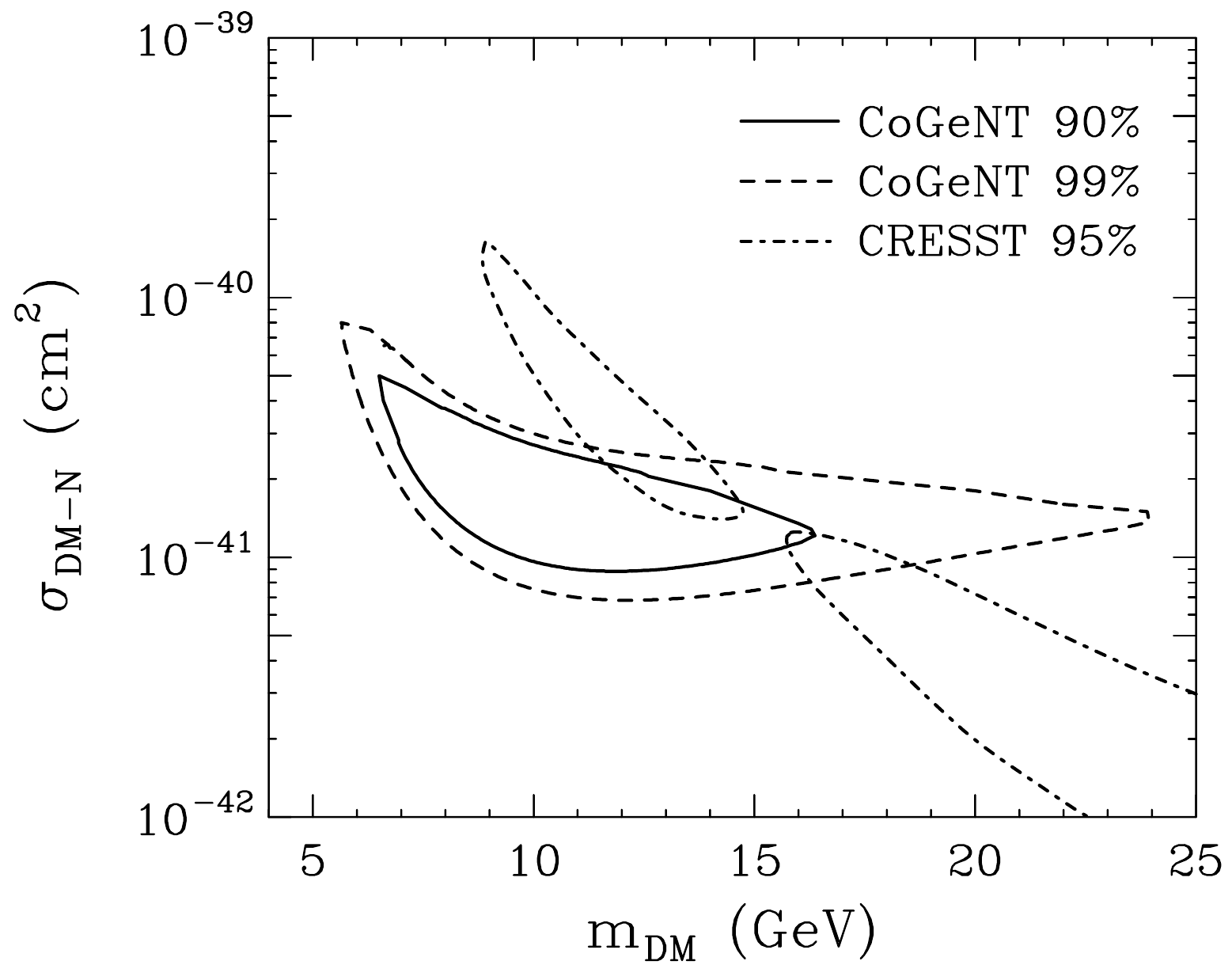}}
\caption{The 90\% (solid) and 99\% (dashed) confidence level contours for the spectrum of events observed by CoGeNT compared to the  95\% confidence level regions favored by CRESST-II (dot-dashed). A dark matter particle with a mass of approximately 10-20 GeV and an elastic scattering cross section with nucleons of approximately $(1-3)\times 10^{-41}$ cm$^2$ can account for the excess events reported by each of these experiments. This figure was adapted from one originally appearing in Ref.~\cite{arXiv:1110.5338}.}
\label{fig:massCross}
\end{figure}

In Fig.~\ref{fig:massCross}, the range of dark matter mass and elastic scattering cross section are shown which can provide a good fit to the spectra of excess low-energy events reported by CoGeNT~\cite{arXiv:1110.5338} and CRESST-II~\cite{Angloher:2011uu}, assuming a Maxwellian velocity distribution with $v_0=220$ km/s and $v_{\rm esc}=544$ km/s (see also Ref.~\cite{kopp}). In fitting to the spectrum of the CoGeNT data, we marginalize over a range of surface event rejection efficiencies, as described in Ref.~\cite{arXiv:1110.5338}. As seen from this figure, the dark matter parameter space favored by CRESST-II is compatible with the region implied by CoGeNT's spectrum. In particular, a dark matter particle with a mass of roughly 10-20 GeV and an elastic scattering cross section with nucleons of $(1-3)\times 10^{-41}$ cm$^2$ could account for the excess events reported by both collaborations.

\subsection{Direct Evidence Summary and Constraints}

In this section, I have summarized the direct detection signals reported by the DAMA/LIBRA, CoGeNT and CRESST-II collaborations. The spectra of excess events observed by CoGeNT and CRESST-II are compatible with arising from the same dark matter particle with a mass of $\sim$10-20 GeV and an elastic scattering cross section with nucleons of $(1-3)\times 10^{-41}$ cm$^2$. Similarly, the modulation amplitudes reported by DAMA/LIBRA and CoGeNT appear to be mutually consistent. Under the standard assumptions of a Maxwellian velocity distribution and velocity-independent scattering cross sections, however, the spectrum and rate of events reported by CoGeNT and CRESST-II would lead one to expect a signficantly smaller (by a factor of about three to five) modulation amplitude than is observed by DAMA/LIBRA and CoGeNT. In order to account for this apparent discrepancy, one can consider dark matter particles 1) with non-Maxwellian features in their velocity distribution, such as local streams~\cite{lightorstream}, or 2) with a velocity-dependent scattering cross section with nuclei~\cite{inelastic,momentumdm,resonant,Fitzpatrick:2010br,Farina:2011pw}, either of which can lead to significant enhancements in the observed modulation amplitude.

High-resolution numerical simulations find that instead of being smooth, the dark matter halos of Milky Way-like galaxies contain many smaller subhalos. Most of these subhalos have a great deal of their outer mass stripped, resulting in the formation of cold tidal streams~\cite{simulations,nonmaxwell}. And while the presence of such streams in the local neighborhood could potentially effect the spectrum of dark matter-induced events observed in direct detection experiments~\cite{nonmaxwelldirect,Lisanti:2010qx}, these effects are often found to be far more pronounced in the modulation signals of such experiments~\cite{Kuhlen:2009vh}.  The presence of such streams can significantly enhance a modulation signal, as well as shift the phase of the modulation relative to that predicted in more simple halo models~\cite{Kuhlen:2009vh,ls,lightorstream}. If $\sim$20-30\% of the local dark matter density is found the form of a velocity stream, the large modulation amplitudes observed by DAMA/LIBRA and CoGeNT could potentially be reconciled with the spectra of excess events reported by CoGeNT and CRESST-II~\cite{arXiv:1110.5338}.



As the CoGeNT collaboration collects more data, the spectrum of the modulation amplitude will become much better measured, making it possible to begin to discriminate between the various options described in this section. By the summer of 2012, the CoGeNT collaboration will have doubled the size of their data set, and plans to deploy the first of four CoGeNT-4 (C4) detectors, roughly quadrupling their effective target mass (the completed C4 experiment will possess a target mass an order of magnitude larger than the existing CoGeNT detector). If streams or resonances are responsible for a significant fracton of the observed modulation, these features will become increasingly apparent as this data set grows.

A number of other direct detection experiments have placed constraints on the elastic scattering cross sections of dark matter particle in the mass range being considered here. I will next review these constraints and discuss their implications for the dark matter interpretation of DAMA/LIBRA, CoGeNT and CRESST-II.

\begin{itemize}

\item{{\bf CDMS Low Threshold Analysis:} The CDMS-II collaboration has presented the results of two analyses searching for low-mass dark matter particles~\cite{Ahmed:2010wy,Akerib:2010pv}. The more stringent of these constraints finds $\sigma \lsim 2.2\times 10^{-41}$ cm$^2$ for a mass of 10 GeV (at the 90\% confidence level). So although this result disfavors the upper range of the elastic scattering cross section capable of accounting for CoGeNT and CRESST-II, an elastic scattering cross section of $\sim(1-2.2)\times 10^{41}$ cm$^2$ is consistent with CoGeNT, CRESST-II and CDMS-II. In fact, the spectrum of low events observed by CDMS is quite similar to that reported by CoGeNT (see Fig.~4 of Ref.~\cite{arXiv:1110.5338}  for a direct comparison). It should be emphasized that if CoGeNT's annual modulation does result from dark matter, then CDMS's low energy events will also demonstrate a considerable degree of annual modulation. Although no CDMS modulation analysis has been presented as of yet, the results of such a study would be very valuable.}

\item{{\bf Constraints from XENON-100 and XENON-10:} The XENON-100~\cite{xenon100} and XENON-10~\cite{xenon10} collaborations have each reported rather strong constraints on the parameter space of low-mass dark matter particles. As presented, these constraints appear to largely rule out the dark matter parameter space collectively favored by CoGeNT and CRESST-II. There are a number of ways, however, in which these constraints could be significantly weaker than they might appear. Firstly, any uncertainties in the response of liquid xenon to very low-energy nuclear recoils (as encapsulated in the functions $L_{\rm eff}$ and/or $Q_y$) could significantly impact the corresponding constraints for dark matter particles with a mass in the range of interest. The constraints from the XENON-100 collaboration were derived using measurements of the scintillation efficiency, $L_{\rm eff}$, as described in Refs.~\cite{plante}, which have been criticized in Ref.~\cite{juan2} (see also Ref.~\cite{Collar:2010ht}). Even modest changes to these values at the lowest measured energies ($\sim$3-4 keV) can lead to much weaker constraints on light dark matter particles. It has also been argued that the relatively large (9.3 eV) band-gap of xenon is expected to lead to a suppression of the response to nuclear recoils in the energy range of interest (see Ref.~\cite{Collar:2010ht} and references therein). Many of these issues also apply to constraints on light dark matter making use of only the ionization signal in liquid xenon detectors~\cite{xenon10}. Alternatively, the constraints from XENON-100 and XENON-10 could be modified if dark matter particles do not have identical couplings to protons and neutrons~\cite{zurek,feng}. In particular, for a ratio of couplings given by $f_n/f_p \approx -0.7$, the constraint from xenon-based experiments is weakened by a factor of $\sim$20 relative to that found in the $f_n=f_p$ case~\cite{feng}. For this ratio of couplings, the cross section favored by CRESST-II would also be moved down by a factor of $\sim$7 relative to that observed by CoGeNT. A ratio of $f_n/f_p \approx -0.6$ would reduce the strength of the XENON-100 and XENON-10 constraints by a factor of 3-4, while also lowering the CRESST-II region (relative to that of CoGeNT) by a similar factor.}

\item{{\bf Constraints from other direct detection experiments:} We briefly mention that although the SIMPLE collaboration has placed constraints on the region of parameter space being considered here~\cite{Felizardo:2011uw}, those results have been strongly criticized in the literature~\cite{Dahl:2011tf,Collar:2011kr}. In particular, it is difficult to reconcile the results of SIMPLE's physics run with its own calibration data~\cite{Dahl:2011tf}. We also note that a constraint based on the CRESST commissioning run data~\cite{Brown:2011dp} appears to be in mild tension with the upper range (in cross section) of the parameter space reported to be favored by the analysis of the CRESST-II collaboration. This result is consistent with the lower range of the parameter space favored by CRESST-II and CoGeNT, however.}

\end{itemize}

\section{Implications for Particle Physics}
\label{particle}

To accommodate the collection of observations summarized in this article, a dark matter candidate must have a number of fairly specific characteristics. In particular, such a particle must possess:
\begin{itemize}
\item{A mass in the range of approximately 7-12 GeV.}
\item{A low-velocity annihilation cross section to each of $e^+ e^-$ and $\tau^+ \tau^-$ of $\sigma v\approx (1-5)\times 10^{-27}$ cm$^3$/s. If we make the not unreasonable assumption that annihilations also proceed to $\mu^+ \mu^-$, we require a total cross section to charged leptons of $\sigma v\approx (3-15) \times 10^{-27}$ cm$^3$/s (or  $\sigma v\sim [1.5-30] \times 10^{-27}$ cm$^3$/s if uncertainties in the dark matter density~\cite{bertonehalo} are taken into account). In addition, up to another approximately 20\% of annihilations could also proceed to hadronic final states.}
\item{A spin-independent elastic scattering cross section with nucleons of approximately $\sigma\approx (1-3)\times 10^{-41}$ cm$^2$.}
\end{itemize}

We could also impose that the dark matter candidate in question be produced in the early universe with a relic density equal to the measured dark matter abundance, which implies that the total annihilation cross section at freeze-out be $\sigma v_{\rm FO} \approx 3 \times 10^{-26}$ cm$^3$/s. Any difference between this value and those required in the second bullet point above could arise from velocity dependent terms in the annihilation cross section, for example. 

In this section, I will summarize the particle physics implications of these observations, and discuss what kind of dark matter particle could account for them (following, in large part, Ref.~\cite{Buckley:2010ve}).

\subsection{Dark Matter's Elastic Scattering Cross Section}

We first consider the requirement of the dark matter's elastic scattering cross section. Such interactions are well suited for an effective field theory approach~\cite{beltran}. In particular, there are relatively few operators we can write down which lead to a sizeable spin-independent scattering cross section in the relevant low-velocity limit. These possibilities consist of elastic scattering mediated by a heavy colored and fractionally charged particle, by a neutral vector boson (the $Z$ or a $Z^{\prime}$), or by a neutral scalar.

In order to generate an elastic scattering cross section as large as required by CoGeNT and CRESST-II ($\sigma \sim 2\times 10^{-41}$ cm$^2$) through an interaction mediated by a colored and fractionally charged particle, $q^{\prime}$, the mass of the mediator must be less than approximately 2 TeV (for perturbative couplings, $g \lsim 1$). Such a state is very likely within the ultimate reach of the LHC. 
 


Alternatively, we can consider dark matter scattering that is mediated by a vector boson, such as the Standard Model $Z$, or a $Z'$. In the case of the Standard Model $Z$, we can obtain the required cross section for small coupling between the dark matter and the $Z$, $g_{Z-{\rm DM}-{\rm DM}}\approx0.007$, well below the constraints from measurements of the invisible $Z$ width ($g_{Z-{\rm DM}-{\rm DM}}\lsim 0.023$) \cite{PDG}. For perturbative couplings, a $Z'$ as heavy as approximately 2.6 TeV could generate the cross section implied by CoGeNT and CRESST-II. And although a heavy $Z'$ with universal couplings to Standard Model fermions is excluded by LEP, a lighter 
($\sim$10 GeV) and thus more weakly coupled $Z'$ need not be leptophobic. A TeV-scale leptophobic $Z'$ with couplings capable of producing the cross section implied by CoGeNT and CRESST-II should quickly become within the reach of the LHC~\cite{Han:2010rf}, although such a state with a mass below $\sim$300 GeV could remain below the sensitivity of the LHC and Tevatron experiments~\cite{Aaltonen:2008dn}. Several models capable of generating the CoGeNT and CRESST-II signals with a $\sim$150 GeV $Z'$ motivated by CDF's recent $W$+dijet excess have been proposed (for example, see Ref.~\cite{Buckley:2011vc}).


Lastly, we can also consider dark matter scattering mediated by a scalar~\cite{Belikov:2010yi}. In particular, a scalar which is a singlet under $SU(2)_L$ can couple directly to the dark matter and to Standard Model quarks through mixing with the Higgs sector. A very light scalar ($m \lsim 10$ GeV) that mixes slightly with the Standard Model Higgs could account for the signals reported by CoGeNT and CRESST-II. If the Higgs sector is more complicated (such as in models with multiple Higgs doublets), heavier scalars could also mediate such an interaction.

If instead of a singlet, the mediating scalar is a doublet under $SU(2)_L$, it can couple directly to quarks. But in this case, the dark matter itself must consist of a mixture of $SU(2)_L$ singlets, doublets, and/or triplets, leading to the introduction of heavy charged states in the dark sector. To evade constraints on charged particles from LEP-II, the dark matter must be primarily singlet, and will posses very small effective couplings to quarks. In such a scenario, the mediating scalar must be lighter than $\sim$20 GeV if cross sections as large as those implied by CoGeNT and CRESST-II are to be generated.


\subsection{Dark Matter Annihilation}

Turning our attention now to dark matter annihilation, we are primarily interested in those interactions which contribute to the annihilation cross section in the low velocity limit and which can primarily result in annihilations to leptons (including $\tau^+ \tau^-$ and $e^+ e^-$). If the dark matter is a Dirac fermion, for example, a leptophilic $Z'$ could mediate such an interaction. To evade constraints from LEP-II, however, such a $Z'$ must be relatively light ($m_{Z'} \lsim 30$ GeV) and somewhat weakly coupled. In principle, the same $Z'$ (with smaller, but non-zero couplings to quarks) could also mediate the elastic scattering cross section observed by CoGeNT and CRESST-II.

The dark matter's annihilations could also be mediated by a scalar, although Yukawa couplings acquired though mixing with the Higgs sector will be unable to provide the necessary annihilations to $e^+ e^-$ (although in some such models, the gamma-ray signal from annihilations to $\tau^+ \tau^-$ can be accommodated~\cite{leptonhiggs}). Alternatively, one could consider annihilations through the $t$-channel exchange of a particle with lepton number (or a scenario in which the dark matter particles which themselves carry lepton number).

\subsection{Asymmetric Dark Matter}


In most models, the abundance of dark matter is simply determined by its self-annihilation cross section and is unrelated to the density of baryons in the universe. From this perspective, it may be somewhat surprising that the cosmological dark matter and baryon densities are of the same order of magnitude, $\rho_{\rm DM}/\rho_b \approx 5$. This observation has motivated models in which the baryon-antibaryon asymmetry that leads to the cosmic baryon abundance is connected to an analogous asymmetry in the dark matter sector~\cite{ADMmodels}. If dark matter particles and quarks each carry the same absolute baryon number, for example, one could account for the observed baryon density without a net baryon number asymmetry if the dark matter were to possess a mass of $m_{\rm DM} = 3 \times (\rho_{\rm DM}/\rho_{b}) \, m_p \approx 14$ GeV. And while the precise ratio of the dark matter and nucleon masses required in such a scenario depends on the details of the operator that transfers the asymmetry between the baryons and dark matter, one generally expects the dark matter to possess a mass on the order of 10 GeV in asymmetric models (for possible exceptions, see Ref.~\cite{randall}). In light of this observation, the evidence for approximately 10 GeV dark matter particles presented here is suggestive of a connection with the baryon asymmetry. 

If the dark matter were to retain such an asymmetry and remain in a pure particle (or pure anti-particle) state indefinately, it would be unable to annihilate and produce the gamma-ray and radio signals discussed in Sec.~\ref{indirect}. In many asymmetric dark matter models, however, particle-antiparticle mixing can efficiently erase such an asymmetry over time, enabling indirect signals to appear~\cite{Buckley:2011ye}.

\subsection{Constraints From Colliders}

Although constraints on relatively light dark matter particles from collider experiments can depend strongly on the details of the particle physics model under consideration, some largely model-independent statements can be made:

\begin{itemize}

\item{{\bf Constraints From LEP:} Light dark matter particles with couplings to electrons can be constrained by monophoton-plus-missing energy searches at LEP. Such constraints are relatively model-independent, and can be made using an effective field theory approach. For a vector, $s$-channel (scalar, $t$-channel) operator with equal couplings to all three generations of charged leptons, Ref.~\cite{Fox:2011fx} finds that LEP data constrains $\sigma v \lsim 2.2 \times 10^{-26}$ cm$^3$/s ($\sigma v \lsim 1.4 \times 10^{-26}$ cm$^3$/s) at the 90\% confidence level. This constraint is a factor of roughly three (two) time weaker than would be required to exclude a dark matter interpretation of the gamma-ray and radio signals described in Sec.~\ref{indirect}, and could be further weakened if the annihilation cross section is mediated by a light particle.}

\item{{\bf Constraints From Hadron Colliders and Prospects For The LHC:}  Much as lepton colliders can constrain dark matter's couplings to electrons by searching for events with a photon and missing energy, the Tevatron and LHC can constrain dark matter's couplings to quarks and gluons by searching for events with missing energy and a single jet~\cite{Beltran:2010ww} . And although current constraints from the Tevatron are still more than two orders of magnitude from the cross sections implied by CoGeNT and CRESST-II~\cite{Bai:2010hh}, future LHC data (operating at 12-14 TeV) should be sensitive to dark matter with the effective couplings required to generate these signals. Again, such constraints could be potentially evaded if the interactions are mediated by light particles~\cite{Fox:2011pm}.}
 
\end{itemize}

\section{Summary and Conclusions}
\label{summary}

In this article, I have attempted to summarize and describe the body of evidence in favor of approximately 10 GeV dark matter particles that has accumulated over the past several years. In my opinion, the case for a dark matter interpretation of this data is very compelling and should be given significant attention and scrutiny. Some of the reasons supporting this opinion include:

\begin{itemize}

\item{Several of the observations described here are very difficult to explain with known or proposed astrophysical backgrounds or systematic effects. Although sources such as pulsars and cosmic ray scattering have been explored to explain the Galactic Center gamma ray flux, they fail to accommodate the observed spectrum and morphology of this signal. The spectra observed from the Milky Way's radio filaments has also been a long standing challenge to explain astrophysically. The annual modulation observed by DAMA/LIBRA (and now supported by CoGeNT) possesses a phase which peaks earlier than any of the possible backgrounds that have been proposed (such as those associated with the atmospheric muon flux and radon decay rate). This has often not been the case in past instances of observations being interpretated in terms of dark matter (such as PAMELA's positron excess~\cite{arXiv:0810.4995}, which could as easily be explained by conventional astrophysics, such as pulsars~\cite{pamelapulsars}, as by dark matter.)}

\item{The collection of observations described here overconstrains the underlying dark matter model in important ways. The gamma-ray, radio filaments, and synchrotron haze signals, for example, each probe the rate and distribution of dark matter annihilations in the Inner Milky Way, and thus are interconnected. In particular, all three of these observations require the same (or very similar) dark matter distribution and annihilation cross section. In this way, these signals are not only consistent with each other, but imply and require each other. Similarly, in order to interpret DAMA/LIBRA's modulation in terms of dark matter, CoGeNT's rate must also demonstrate a degree of variation comparable to that observed. The spectrum of excess low-energy events reported by CoGeNT and CRESST-II are also compatible and imply similar values for the dark matter's mass and elastic scattering cross section with nuclei.}

\item{The characteristics of the dark matter particle and its distribution implied by these observations is consistent with conventional theoretical expectations. In particular, the normalization of the gamma-ray, radio filaments, and synchrotron haze signals each require an annihilation cross section to leptons that is similar (within a factor of a few) of the value predicted for a simple thermal relic ($\sigma v \approx 3 \times 10^{-26}$ cm$^3$/s). No boost factors or other enhancements are required. Furthermore, the dark matter distribution that is required to accommodate these signals ($\rho_{\rm DM} \propto r^{-1.3}$) is highly consistent with the predictions of state-of-the-art hydrodynamical simulations, as well as with observations.}

\end{itemize}

I have reviewed constraints on this dark matter scenario from gamma-ray studies of dwarf spheroidal galaxies, distortions of the cosmic microwave background, the positron and antiproton cosmic ray spectra, energetic neutrinos from the Sun, CDMS, LEP, the Tevatron, and the LHC. In each case, I have found consistency with the dark matter interpretation being put forth here. Consistency with the results of the XENON-10 and XENON-100 collaborations requires either a suppression in the response of liquid xenon to low-energy nuclear recoils, or destructive inference between the dark matter's couplings to protons and neutrons. In the relatively near future, data from Planck, AMS-02, CDMS, and the LHC could be able to further strengthen (or weaken) the case for a dark matter interpretation of these signals.

\bigskip 


\bigskip
\bigskip
\bigskip

{\it Acknowledgements}: In preparing this article, I have relied heavily on previous work by Matt Buckley, Lisa Goodenough, Chris Kelso, Tim Linden, Tim Tait, and others. I would also like to thank Matt Buckley, Juan Collar, Paddy Fox, Tim Linden, and Ethan Neil for many helpful comments and discussions. This work has been supported by the US Department of Energy and by NASA grant NAG5-10842.

\end{document}